\documentclass[preprint,showpacs,showkeys]{revtex4-1}
\usepackage{graphicx}
\usepackage[caption=false]{subfig}
\usepackage{graphicx,amsmath}
\usepackage{epsfig}
\usepackage{amssymb}
\usepackage{amsmath}
\usepackage{lipsum}
\usepackage{subfig}
\usepackage{mathrsfs}
\RequirePackage[colorlinks,citecolor=blue,urlcolor=magenta,linkcolor=magenta]{hyperref}
\begin{document} 
\title{Froissart bound and self-similarity based models of proton structure functions}
\author{D. K. Choudhury$^1$}
\author{Baishali Saikia$^1$}
\email[Corresponding author : \ ]{baishalipiks@gmail.com}
\affiliation{$^1$Department of Physics, Gauhati University , Guwahati- 781 014, Assam, India} 
\begin{abstract}
Froissart Bound implies that the total proton-proton cross-section (or equivalently structure function) cannot rise faster than the logarithmic growth $\log^2 s \sim \log^2 1/x$, where \textit{s} is the square of the center of mass energy and \textit{x} is the Bjorken variable. Compatibility of such behavior with the notion of self-similarity in a model of structure function suggested by us sometime back is now generalized to more recent improved self-similarity based models and compare with recent data as well as with the model of Block, Durand, Ha and McKay. Our analysis suggests that Froissart bound compatible self-similarity based models are possible with $\log^2 1/x$ rise in limited  $x-Q^2$ ranges of HERA data, but their phenomenological ranges validity are narrower than the corresponding models having power law rise in $1/x$. 
\textbf{(Dedicated to Professor D. P. Roy (1941-2017) )}
\end{abstract}
\maketitle

\section{Introduction}
One of the cornerstones of the present strong interaction physics is the Froissart theorem \cite{fe}. It declares that the total cross section of any two-hadron scattering cannot grow with energy faster than $(\log s)^2$ where \textit{s} is the center of mass energy square. Later it was improved by Martin \cite{mrt2,mrtn,yjp}. The original derivation of Froissart \cite{fe} is based on Mandelstam representation and that of Martin \cite{mrt2,luu} is on axiomatic field theory which could be considered as more general. The approach has led further development of the subject \cite{roy,royk,mrt3, roy1, roy3} as well as construction of several phenomenological models \cite{phm1,phm2}. It is therefore as familiar as Froissart-Martin bound. 

Precession measurement of proton-proton (\textit{pp}) cross-section at LHC \cite{lhc1,lhc2,lhc3,lhc4} and in cosmic rays \cite{cosmi} have led the PDG group \cite{pdg} to fit the data with such $\log^2 s$ term together with an additional constant $\sigma \sim A +B \log ^2 s$. There is  also an alternative fit for pp data \cite{roh1} with an addition of non leading $\log s$ term

Exact proof of Froissart Saturation in QCD is not yet been reported. However, in specific models, such behavior is found to be realizable. Specifically, soft gluon resummation models in the infrared limit of QCD \cite{roh2} and /or gluon-gluon recombination as in GLR \cite{glrr} equation or color glass condensate \cite{roh3, roh4, roh5} models  such $\log^2 s$ rise of proton proton cross section is achievable.

In DIS, when Froissart bound is related to the nucleon structure function $F_2(x,Q^2)$, it implies a growth limited to $\log^2\dfrac{1}{x}$.

It is well known that the conventional equations of QCD, like DGLAP \cite{o2,o3,o1} and BFKL approaches \cite{o6,o4,o5,o7}, this limit is violated; while in the DGLAP approach, the small-\textit{x} gluons grow faster than any power of $\ln \left( \dfrac{1}{x}\right)  \approx \ln \left(\dfrac{s}{Q^2} \right)$ \cite{rg}, in the BFKL approach it grows as a power of $\left( \dfrac{1}{x}\right) $ \cite{o6,o4,o5,o7,o8}.

However, in recent years, the validity of Froissart Bound for the structure function at phenomenological level has attracted considerable attention in the study of DIS, mostly due to the efforts of Block and his collaborators \cite{bff,blooo,blo,buu,bu}. 

It was argued in Ref. \cite{buu} that as the structure function $F_2^{\gamma p}(x,Q^2)$ is essentially the total cross section for the scattering of an off-shell gauge boson $\gamma^*$ on the proton, a strong interaction process up to the initial and final gauge boson-quark couplings and Froissart bound makes sense. On this basis, one analytical expression in \textit{x} and $Q^2$ for the DIS structure function has been suggested by them \cite{blooo} which has expected Froissart compatible $\log^2 \dfrac{1}{x}$ behavior and valid within the range of $Q^2$: $0.85\leq Q^2 \leq 1200$ GeV$^2$ of the HERA data. Using this expression as input at $Q_0^2 = 4.5$ GeV$^2$ to DGLAP evolution equation, the validity is increased upto 3000 GeV$^2$ \cite{blo}. The approach has been more recently applied in the Ultra High Energy (UHE) neutrino interaction, valid upto ultra small \textit{x} $\sim 10^{-14}$ \cite{bu} . It is therefore of interest to study if such Froissart saturation like behavior can be incorporated in any other proton structure functions as well and can be tested with data.

The aim of the present paper is exactly this: we will study the possibility of incorporating Froissart saturation like behaviour in the parametrization of structure function of nucleon based on self-similarity as suggested by Lastovicka \cite{Last} and later pursued by us \cite{dkc,DK4,DK5,DK6,bs1,bsc}. Specifically in Ref \cite{dkc}, such possibility was first suggested. The present work is a generalization and improvement of it in the sense that the improved models can incorporate linear rise in $\log Q^2$ instead of $Q^2$ and make them closer to data and QCD expectation. As the physics of small \textit{x} is not yet fully understood, it is a worthwhile study, which  needs to be  tested with most recent data. This is  the  aim  of the present paper.

In section \ref{forma}, we will show the required formalism while in section \ref{summ}, summary of the present work will be discussed.

\section{Formalism}
\label{forma}
\subsection{Froissart bound in self-similarity based Proton structure function}

The possibility of incorporating Froissart bound in the self-similarity based model of proton structure function suggested by Lastovicka \cite{Last} was first attempted in Ref. \cite{dkc}. In the model of Ref \cite{Last}, the magnification factors were defined as $M_1 = \left(1 +\frac{Q^2}{Q_0^2} \right) $ and $M_2= \dfrac{1}{x}$. It was noted in \cite{dkc} that if the magnification factor $M_2$ is changed to $\left( \log \dfrac{1}{x}\right) $, then it is possible to get a Froissart saturation like behavior in structure function. However, we observe that it is true only for PDF but not for the structure function. 

Below we address this point. For completeness, we first outline the self-similarity based model of proton structure function of Ref. \cite{Last}

The self-similarity based model of the proton structure function of Ref\cite{Last} is based on parton distribution function(PDF) $q_i(x,Q^2)$. Choosing the magnification factors $M_1= \left(1+\dfrac{Q^2}{Q_0^2}\right)$ and $M_2= \left(\dfrac{1}{x}\right)$, the unintegrated Parton Density (uPDF) can be written as \cite{Last,DK6} 
\begin{equation}
\label{E1}
\log[M^2.f_i(x,Q^2)]= D_1.\log\dfrac{1}{x}.\log\left(1+\dfrac{Q^2}{Q_0^2}\right)+D_2.\log\dfrac{1}{x}+D_3.\log\left(1+\dfrac{Q^2}{Q_0^2}\right)+D_0^i
\end{equation}
\\
where \textit{x} is the Bjorken variable and $Q^2$ is the renormalization scale and \textit{i} denotes a quark flavor. Here $D_1,\ D_2,\ D_3$ are the three flavor independent model parameters while $D_0^i$ is the only flavor dependent normalization constant. $M^2$ is introduced to make (PDF) $q_i(x,Q^2)$ as defined below (in Eqn \ref{E2}) dimensionless which is set to be  as 1 GeV$^2$ \cite{DK6}. We note that in deriving the model ansatz Eqn (\ref{E1}), one has to first generalize the definition of dimension from discrete to continuous fractals. The proper choice of magnification factors are made on the condition that they should be positive, non-zero and have no physical dimension. Whereas, in Ref\cite{Last}, choice of $\left( 1+\dfrac{Q^2}{Q_0^2}\right)$ is made and an equivalent choice of $\left( \dfrac{Q_0^2}{Q^2+Q_0^2}\right)$ is also equally plausible. So is $\left( \dfrac{1}{x}\right)$ vs $\left( \log \dfrac{1}{x}\right)$. The integrated quark densities (PDF) then can be defined as
\begin{equation}
\label{E2}
q_i(x,Q^2) = \int_0^{Q^2}f_i(x,Q^2)dQ^2
\end{equation}
\\
As a result, the following analytical parametrization of a quark density is obtained by using Eqn(\ref{E2}) \cite{DK5}\\
\begin{equation}
\label{E3}
q_i(x,Q^2) = e^{D_0^i}f(x,Q^2)
\end{equation}
where
\begin{equation}
\label{E4}
f(x,Q^2)= \frac{Q_0^2 \ \left( \frac{1}{x}\right) ^{D_2}}{M^2\left(1+D_3+D_1\log\left(\frac{1}{x}\right)\right)} \left(\left(\frac{1}{x}\right)^{D_1\log \left(1+\frac{Q^2}{Q_0^2}\right)} \left(1+\frac{Q^2}{Q_0^2}\right)^{D_3+1}-1 \right)
\end{equation}
\\
is flavor independent. Using Eqn(\ref{E3}) in the usual definition of the structure function $F_2(x,Q^2)$, one can get
\begin{equation}
\label{E5}
F_2(x,Q^2)=x\sum_i e_i^2 \left( q_i(x,Q^2)+ \bar{q}_i(x,Q^2)\right) 
\end{equation}
or it can be written as
\begin{equation}
\label{E6}
F_2(x,Q^2)=e^{D_0}xf(x,Q^2)
\end{equation}
\\
where 
\begin{equation}
\label{Ea}
e^{{D_0}}=\sum_{i=1}^{n_f}e^{2}_{i}\left(e^{D_0^i}+ e^{\bar{D}_0^i}\right)
\end{equation}
\\
Eqn(\ref{E5}) involves both quarks and anti-quarks. As in Ref\cite{Last} we use the same parametrization both for quarks and anti-quarks. Assuming the quark and anti-quark have equal normalization constants, we obtain for a specific flavor
\begin{equation}
\label{Eb}
e^{{D_0}}=\sum_{i=1}^{n_f}e^{2}_{i}\left(2 e^{D_0^i}\right)
\end{equation}
\\
It shows that the value of $D_0$ will increase as more and more number of flavors contribute to the structure function.\\ With $n_f=3 , 4 $ and 5 it reads explicitly as
\begin{eqnarray}
\label{Ec}
n_f=3 &:& \ \ e^{D_0}= 2 \left( \frac{4}{9}e^{{D_0}^{u}}+\frac{1}{9}e^{{D_0}^{d}}+\frac{1}{9}e^{{D_0}^{s}} \right) \\
\label{Ed}
n_f=4 &:& \ \ e^{D_0}= 2 \left(\frac{4}{9}e^{{D_0}^{u}}+\frac{1}{9}e^{{D_0}^{d}}+\frac{1}{9}e^{{D_0}^{s}}+\frac{4}{9}e^{{D_0}^{c}}\right) \\
\label{Ef}
n_f=5 &:& \ \ e^{D_0}= 2 \left(\frac{4}{9}e^{{D_0}^{u}}+\frac{1}{9}e^{{D_0}^{d}}+\frac{1}{9}e^{{D_0}^{s}}+\frac{4}{9}e^{{D_0}^{c}}+\frac{1}{9}e^{{D_0}^{b}}\right) 
\end{eqnarray}
\\
Since each term of right hand sides of Eqn(\ref{Ec}),(\ref{Ed}), and (\ref{Ef}) is positive definite, it is clear, the measured value of $D_0$ increases as $n_f$ increases. However, single determined parameter $D_0$ can not ascertain the individual contribution from various flavors.
\\ \\
From HERA data \cite{H1,ZE}, Eqn(\ref{E6}) was fitted in Ref\cite{Last} with
\begin{eqnarray}
\label{E7}
D_0 &=& 0.339\pm 0.145 \nonumber \\
D_1 &=& 0.073\pm 0.001 \nonumber \\
D_2 &=& 1.013\pm 0.01 \nonumber \\
D_3 &=& -1.287\pm 0.01 \nonumber \\
Q_0^2 &=& 0.062\pm 0.01 \ {\text G\text e\text V^2}
\end{eqnarray}

in the kinematical region,
\begin{eqnarray}
\label{E8}
& & 6.2\times10^{-7}\leq x\leq 10^{-2} \nonumber \\
& & 0.045\leq Q^2 \leq 120 \ {\text G\text e\text V^2}
\end{eqnarray}
Following the method of Ref. \cite{dkc} for very small \textit{x} and large $Q^2$, we can write the PDF as
\\
\begin{equation}
\label{qc}
q_i(x,Q^2)=\frac{e^{\acute{D}_0^i}\ \acute{Q}_0^2\ \left( \log \frac{1}{x}\right) ^{\acute{D}_2+\acute{D}_1\log\left(1+\frac{Q^2}{\acute{Q}_0^2}\right)}}{M^2\left(1+\acute{D}_3+\acute{D}_1\log(\log 1/x)\right)} \left(1+\frac{Q^2}{\acute{Q}_0^2}\right)^{\acute{D}_3+1}
\end{equation}
\\
and the corresponding structure function as
\begin{equation}
\label{qd}
\acute{F}_2(x,Q^2)=\frac{e^{\acute{D}_0}\ \acute{Q}_0^2\ x\ \left( \log \frac{1}{x}\right) ^{\acute{D}_2+\acute{D}_1\log\left(1+\frac{Q^2}{\acute{Q}_0^2}\right)}}{M^2\left(1+\acute{D}_3+\acute{D}_1\log(\log 1/x)\right)} \left(1+\frac{Q^2}{\acute{Q}_0^2}\right)^{\acute{D}_3+1}
\end{equation}
If an extra condition on the exponent of $\left( \log\dfrac{1}{x} \right)$ of Eq. \ref{qd} i.e. $\acute{D}_2+\acute{D}_1\log\left(1+\frac{Q^2}{\acute{Q}_0^2}\right)=2$ is imposed, then PDF of Eq. \ref{qc} shows Froissart saturation behavior $\sim \left( \log\dfrac{1}{x} \right)^2$. But it is not so for the structure function of Eq. \ref{qd} due to the additional multiplicative factor \textit{x}. 

If we recast the multiplicative factor \textit{x} of Eq. \ref{qd} as
\begin{equation}
x=\left(\log \frac{1}{x} \right)^A
\end{equation}
with
\begin{equation}
A= \dfrac{-\log\frac{1}{x}}{\log\left(\log \frac{1}{x} \right)}
\end{equation}
then the Froissart condition on the structure function (Eq. \ref{qd}) will be
\begin{equation}
\label{fc}
\dfrac{-\log\frac{1}{x}}{\log\left(\log \frac{1}{x} \right)}+ \acute{D}_2+\acute{D}_1\log\left(1+\frac{Q^2}{\acute{Q}_0^2}\right)=2
\end{equation}
The first term in LHS of Eq. \ref{fc} is negative for $0<x<1$ and independent of the model parameters. For very small $D_2, D_1\sim 0$ the condition of Eq. \ref{fc} will be invalid and hence the general Froissart saturation like behavior in structure function is not possible in the model of Ref. \cite{Last}. Therefore we choose an alternative way to get a proper Froissart Bound condition.

\subsection{Froissart bound compatible self-similarity based Proton structure functions with three magnification factors and power law rise in $Q^2$}
\subsection*{{\large Model 1}}
Taking three magnification factors instead of two:
\begin{eqnarray}
M_1 &=& \left(1+\frac{Q^2}{Q_0^2}\right) \nonumber \\
M_2 &=& \dfrac{1}{x} \nonumber \\
M_3 &=& \log \frac{1}{x}
\end{eqnarray}
one can construct uPDF, PDF and structure function as: \\ \\
uPDF
\begin{multline}
\log[M^2.\grave{f}_i(x,Q^2)]=\grave{D}_1 \log M_1 \log M_2 \log M_3 + \grave{D}_2 \log M_1 \log M_2 + \grave{D}_3 \log M_2 \log M_3 
\\
+ \grave{D}_4 \log M_1 \log M_3 + \grave{D}_5 \log M_1 + \grave{D}_6 \log M_2 + \grave{D}_7 \log M_3 + \grave{D}_0{^i}
\end{multline}
leads to
\begin{multline}
\grave{f}_i(x,Q^2)= e^{\grave{D}_0^i} \ \left(\dfrac{1}{x} \right) ^{\grave{D}_2 \log\left(1+\frac{Q^2}{\grave{Q}_0^2}\right)+\grave{D}_6} 
\\
\times \left(\log \dfrac{1}{x} \right) ^{\grave{D}_1 \log\left(1+\frac{Q^2}{\grave{Q}_0^2}\right) \log 1/x + \grave{D}_3 \log 1/x + \grave{D}_4\log\left(1+\frac{Q^2}{\grave{Q}_0^2}\right) + \grave{D}_7 }\left(1+\frac{Q^2}{\grave{Q}_0^2}\right)^{\grave{D}_5}
\end{multline}
\\
and the corresponding PDF
\begin{multline}
\label{x1}
\grave{q}_i(x,Q^2)= \frac{e^{\grave{D}_0^i}\ \grave{Q}_0^2\ (1/x)^{\grave{D}_6}\ \left( \log \frac{1}{x}\right) ^{\grave{D}_3 \log \frac{1}{x} + \grave{D}_7}}{M^2\left(1+\grave{D}_5+\grave{D}_2\log \frac{1}{x} +(\grave{D}_4+\grave{D}_1\log \frac{1}{x})\log \log \frac{1}{x}\right)} 
\\ \\
\times \left( (1/x)^{\grave{D}_2 \log\left(1+\frac{Q^2}{\grave{Q}_0^2}\right)} (\log 1/x)^{\log\left(1+\frac{Q^2}{\grave{Q}_0^2}\right)\left(\grave{D}_4+\grave{D}_1 \log \frac{1}{x}\right) }\left(1+\frac{Q^2}{\grave{Q}_0^2}\right)^{\grave{D}_5+1} -1\right) 
\end{multline}
For very small \textit{x} and large $Q^2$ , the second term of Eq. (\ref{x1}) can be neglected, leading to

\begin{multline}
\label{x2}
\grave{q}_i(x,Q^2)= \frac{e^{\grave{D}_0^i}\ \grave{Q}_0^2\ (1/x)^{\grave{D}_2 \log\left(1+\frac{Q^2}{\grave{Q}_0^2}\right)+ \grave{D}_6}}{M^2\left(1+\grave{D}_5+\grave{D}_2\log \frac{1}{x} +(\grave{D}_4+\grave{D}_1\log \frac{1}{x})\log \log \frac{1}{x}\right)}
\\ \\
\times\left( \log \frac{1}{x}\right)^{\grave{D}_7 + \grave{D}_3 \log \frac{1}{x} + \left( \grave{D}_4+\grave{D}_1 \log \frac{1}{x}\right)\times \log\left(1+\frac{Q^2}{\grave{Q}_0^2}\right)} \ \left(1+\frac{Q^2}{\grave{Q}_0^2}\right)^{\grave{D}_5+1} 
\end{multline}
from which one can define structure function as:
\begin{multline}
\label{x3}
\grave{F}_2(x,Q^2)= \frac{e^{\grave{D}_0}\ \grave{Q}_0^2\ (1/x)^{\grave{D}_2 \log\left(1+\frac{Q^2}{\grave{Q}_0^2}\right)+ \grave{D}_6-1}}{M^2\left(1+\grave{D}_5+\grave{D}_2\log \frac{1}{x} +(\grave{D}_4+\grave{D}_1\log \frac{1}{x})\log \log \frac{1}{x}\right)} 
\\ \\
\times\left( \log \frac{1}{x}\right) ^{\grave{D}_7 + \grave{D}_3 \log \frac{1}{x} + \left( \grave{D}_4+\grave{D}_1 \log \frac{1}{x}\right)\times \log\left(1+\frac{Q^2}{\grave{Q}_0^2}\right)} \ \left(1+\frac{Q^2}{\grave{Q}_0^2}\right)^{\grave{D}_5+1} 
\end{multline}
which has total 9 parameters: $\grave{Q}_0^2$ and $\grave{D}_i$s with $i=$ 0 to 7. \\ \\
Eq. \ref{x3} can show the proper Froissart saturation behavior in the structure function under the following conditions on the model parameters:
\begin{eqnarray}
\label{lab}
& (1) & \grave{D}_2 \log\left(1+\frac{Q^2}{\grave{Q}_0^2}\right)+ \grave{D}_6 = 1  \nonumber \\ 
& (2) & \grave{D}_7 + \grave{D}_3 \log \frac{1}{x} + \left( \grave{D}_4+\grave{D}_1 \log \frac{1}{x}\right)\times \log\left(1+\frac{Q^2}{\grave{Q}_0^2}\right) = 2 
\end{eqnarray}
Further if $\grave{D}_7, \ \grave{D}_3, \ \grave{D}_1\ll \grave{D}_4$ in \ref{lab}, then $\grave{D}_4=\dfrac{2-\grave{D}_7}{\log\left(1+\frac{Q^2}{\grave{Q}_0^2}\right)}$, the Froissart compatible structure function will be
\begin{equation}
\label{x4}
\grave{F}_2(x,Q^2)= \frac{e^{\grave{D}_0}\ \grave{Q}_0^2\ \left( \log \frac{1}{x}\right) ^2 \ \left(1+\frac{Q^2}{\grave{Q}_0^2}\right)^{\grave{D}_5+1}}{M^2\left(1+\grave{D}_5+\grave{D}_2\log \frac{1}{x} +(\grave{D}_4+\grave{D}_1\log \frac{1}{x})\log \log \frac{1}{x}\right)} 
\end{equation}
\\
which reduces the number parameters by 3. So Eq. \ref{x4} indicates that a self-similarity based model is compatible with Froissart bound having a power law growth in $Q^2$.\\

Using HERAPDF1.0 \cite{HERA}, Eq. \ref{x4} is fitted and found its phenomenological ranges of validity: $1.3\times 10^{-4}\leq x \leq 0.02$ and $6.5 \leq Q^2 \leq 90$ GeV$^2$ with the fitted parameters listed in Table \ref{c7t1}. 

\begin{figure}[tbp]
\captionsetup[subfigure]{labelformat=empty}
\centering
  \subfloat[]{\includegraphics[width=.3\textwidth]{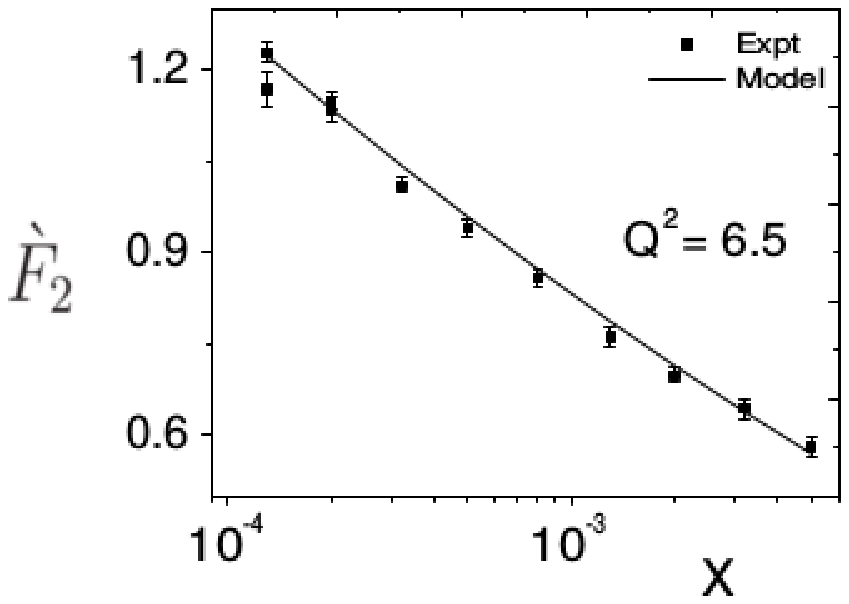}}\quad
 \subfloat[]{\includegraphics[width=.29\textwidth]{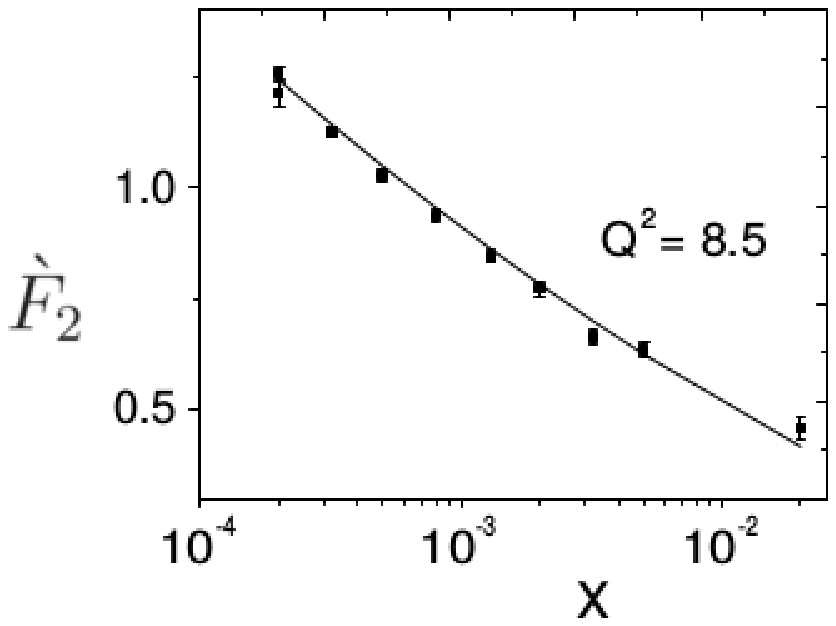}}\quad
\subfloat[]{\includegraphics[width=.3\textwidth]{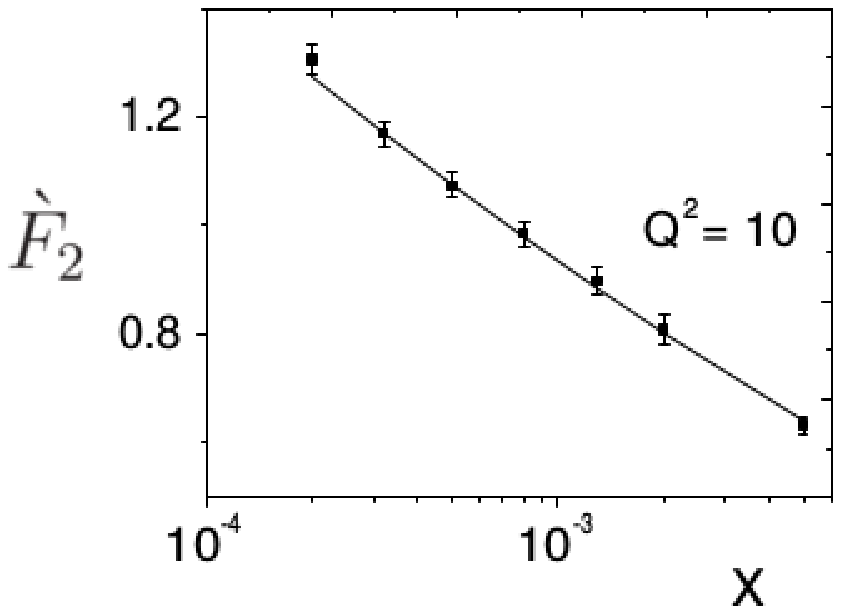}}\quad
 \subfloat[]{\includegraphics[width=.3\textwidth]{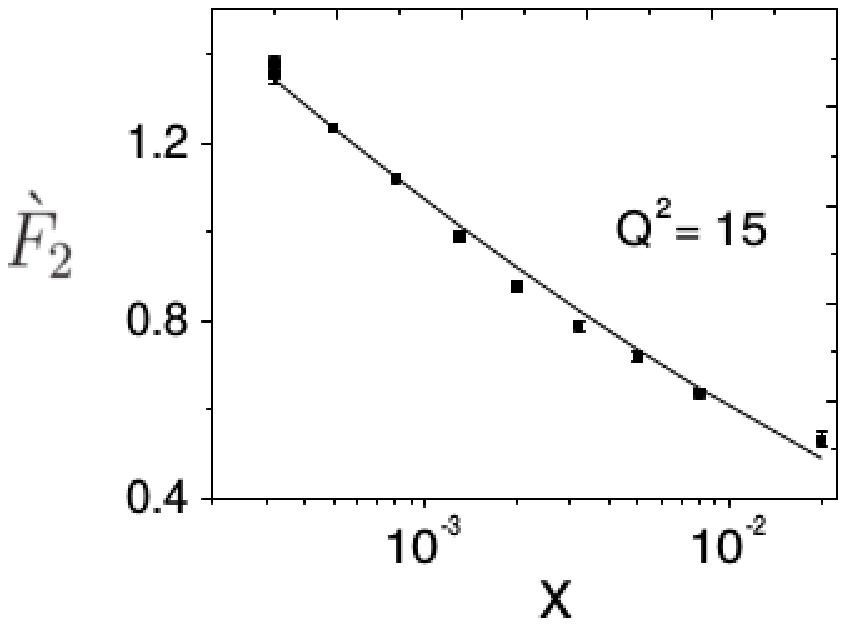}}\quad
 \subfloat[]{\includegraphics[width=.3\textwidth]{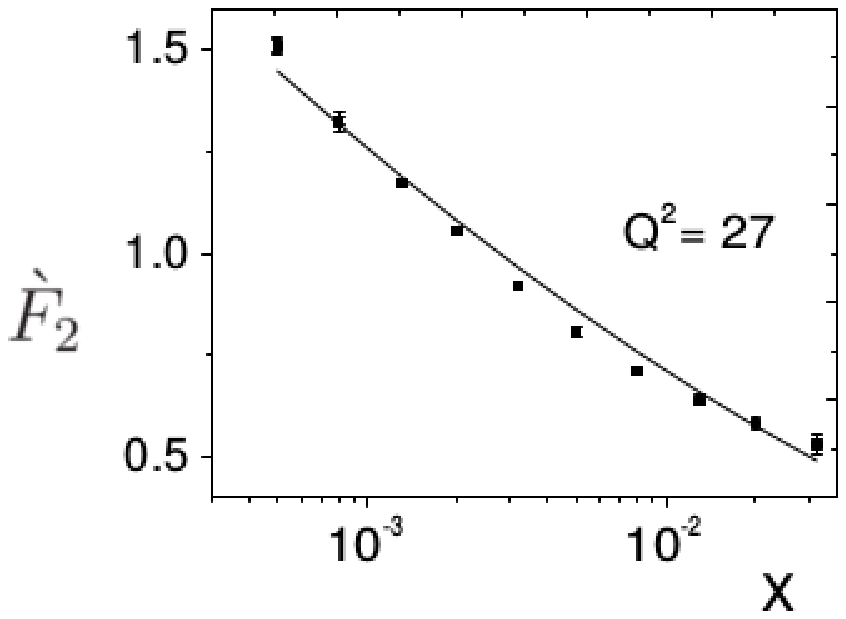}}\quad
  \subfloat[]{\includegraphics[width=.3\textwidth]{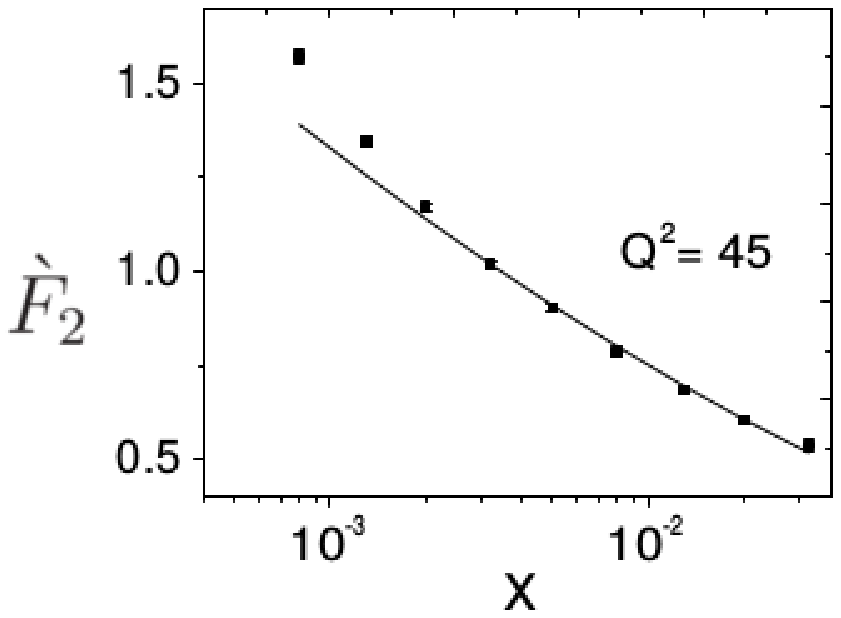}}\quad
\subfloat[]{\includegraphics[width=.31\textwidth]{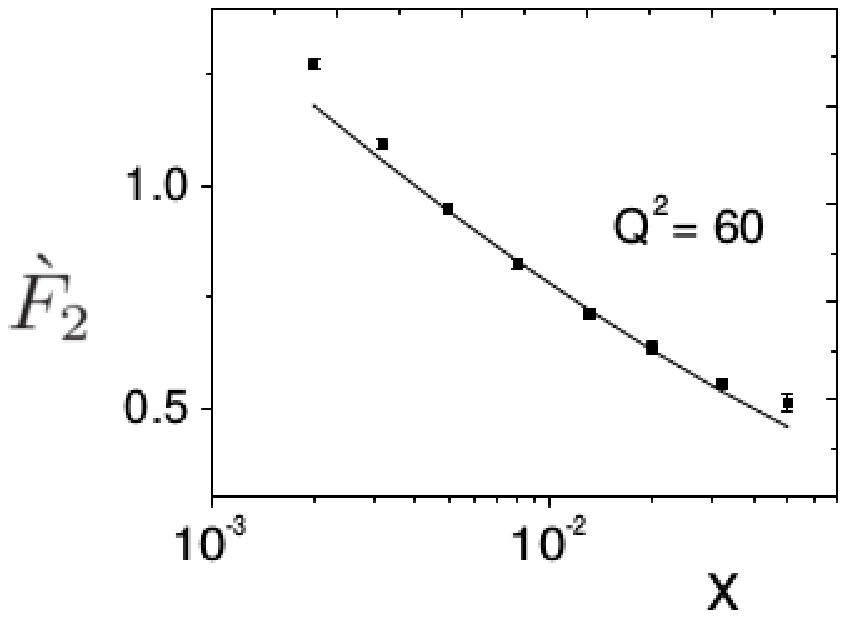}}\quad
\subfloat[]{\includegraphics[width=.3\textwidth]{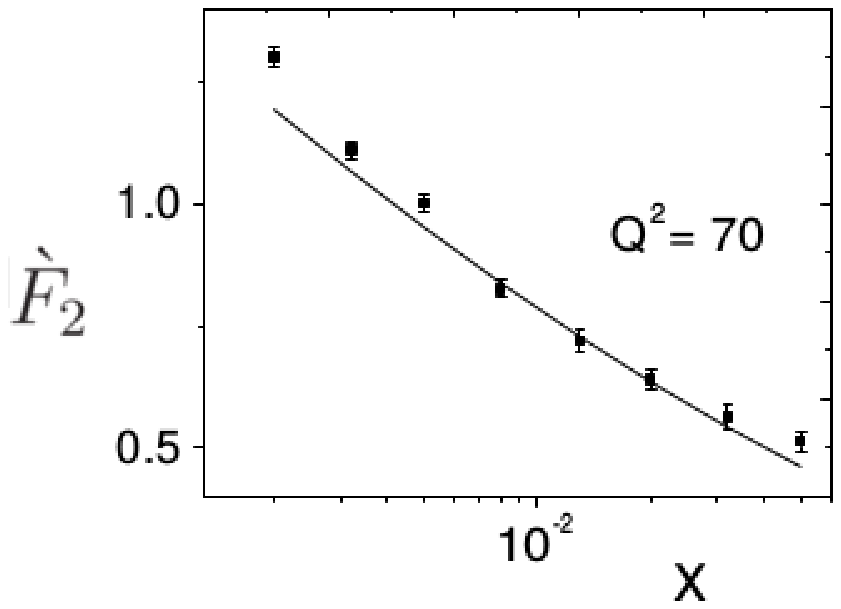}}\quad
 \subfloat[]{\includegraphics[width=.3\textwidth]{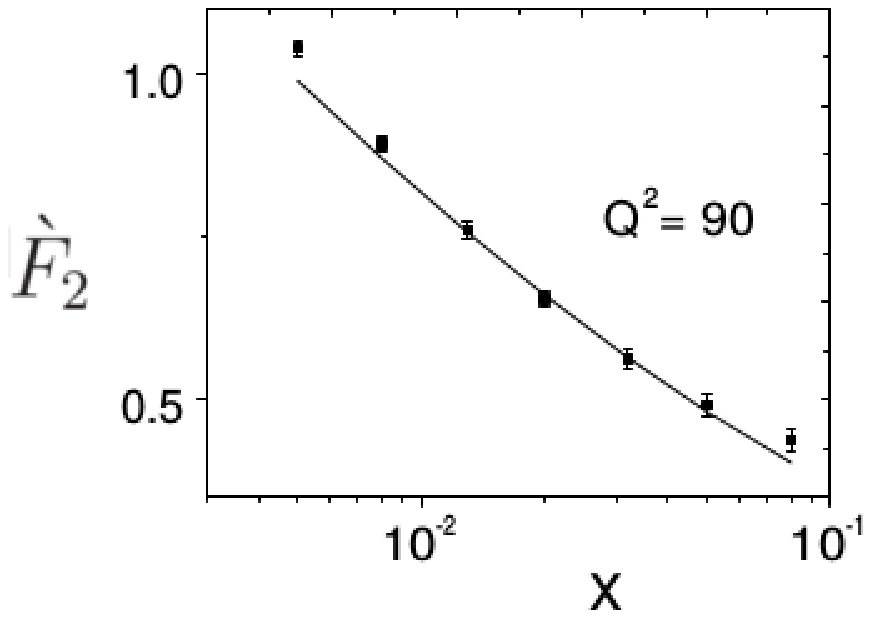}}\quad
 \caption{Comparison of the structure function $\grave{F}_2$ (Eq \ref{x4}; model 1) as a function of $x$ in bins of $Q^2$ with measured data of $F_2$ from HERAPDF1.0 \cite{HERA}}
 \label{c7F1}
\end{figure}

In Fig. \ref{c7F1}, we have shown the graphical representation of $\grave{F}_2$ with data for a few representative values of $Q^2$.
\begin{table}[!tbp]
\caption{\label{c7t1}%
\textit{Results of the fit of} $\grave{F}_2$, model 1, Eq.\ref{x4}}
\begin{ruledtabular}
\begin{tabular}{ccccccc}
\textrm{$\grave{D}_0$}&
\textrm{$\grave{D}_1$}&
\textrm{$\grave{D}_2$}&
\textrm{$\grave{D}_4$}&
\textrm{$\grave{D}_5$}&
\textrm{$Q_0''^2$(GeV$^2$)}&
\textrm{$\chi^2$/ndf} \\
\colrule
0.1006\tiny${\pm 0.003}$ & 0.028\tiny${\pm 0.0008}$ & -0.036\tiny${\pm 0.0001}$ & 3.585\tiny${\pm 0.05}$ & -0.857\tiny${\pm 0.01}$ & 0.060\tiny${\pm 0.001}$ & 0.11 \\
\end{tabular}
\end{ruledtabular}
\end{table}

\subsection*{{\large Model 2}}
The above observation of necessity of having 3 magnification factors can be applied to improve self-similarity based models suggested in Ref. \cite{bs1}. In this case, we can construct another new set of magnification factors:
\begin{eqnarray}
\hat{M}_1 &=& \sum_{j=1}^{n} \frac{B_j}{\left( 1+\frac{Q^2}{\hat{Q}_0^2}\right) ^j}  \quad \quad j=1,2 \nonumber \\
M_2 &=& \dfrac{1}{x} \nonumber \\
M_3 &=& \log \frac{1}{x}
\end{eqnarray}
Here the magnification factor $M_1$ can be considered as special case of a more general form :
\begin{equation}
\label{E30}
\hat{M}_1=\sum_{i=-n}^{n} \alpha_i M_1^i
\end{equation}
Only in a specific case, where $\alpha _1=1$ and all other coefficients cases vanish lead to the original $M_1$ as defined in Eq. \ref{E1}. If we take this generalization form of Eq. \ref{E30} and if all the coefficients $\alpha _i (i=0, \ 1,\ 2, \ . \ . \ . \ , n)$ vanish then Eq. \ref{E30} becomes

\begin{equation}
\label{E33}
\hat{M}_1=\sum_{j=1}^{n} \frac{B_j}{\left( 1+\frac{Q^2}{\hat{Q}_0^2}\right) ^j}
\end{equation}
\\
where
\begin{equation}
B_j= \alpha_{-j}
\end{equation}
as discussed in Ref. \cite{bs1}. Then we can define uPDF, PDF and structure function as follows:\\ \\
The defining equation of uPDF:
\begin{multline}
\log[M^2.\ddot{f}_i(x,Q^2)]= \ddot{D}_1 \log \hat{M}_1 \log M_2 \log M_3 + \ddot{D}_2 \log \hat{M}_1 \log M_2 + \ddot{D}_3 \log M_2 \log M_3 
\\
+ \ddot{D}_4 \log \hat{M}_1 \log M_3 + \ddot{D}_5 \log \hat{M}_1 + \ddot{D}_6 \log M_2 + \ddot{D}_7 \log M_3 + \ddot{D}_0{^i}
\end{multline}
leads to
\begin{equation}
\ddot{f}_i(x,Q^2)= e^{\ddot{D}_0^i}\ \ddot{Q}_0^2\ \left(\dfrac{1}{x} \right) ^{\ddot{D}_6} \left( \log \frac{1}{x} \right) ^{\ddot{D}_3 \log \frac{1}{x}+\ddot{D}_7} \ddot{B}_1 \left[ \frac{1}{\left( 1+ \frac{Q^2}{\ddot{Q}_0^2}\right)} +\frac{\ddot{B}_2}{\ddot{B}_1} \frac{1}{\left( 1+ \frac{Q^2}{\ddot{Q}_0^2}\right)^2}\right]  
\end{equation}
The corresponding PDF and structure function will have the forms
\begin{equation}
\ddot{q}_i(x,Q^2)= e^{\ddot{D}_0^i}\ \ddot{Q}_0^2\ (1/x)^{\ddot{D}_6} \left( \log \frac{1}{x} \right) ^{\ddot{D}_3 \log \frac{1}{x}+\ddot{D}_7} \ddot{B}_1 \left[ \log \left( 1+ \frac{Q^2}{\ddot{Q}_0^2}\right) - \frac{\ddot{B}_2}{\ddot{B}_1} \left( \frac{1}{\left( 1+\frac{Q^2}{\ddot{Q}_0^2}\right)}-1\right) \right]  
\end{equation}
and
\begin{multline}
\label{ddot}
\ddot{F}_2(x,Q^2)= e^{\ddot{D}_0}\ \ddot{Q}_0^2\ (1/x)^{\ddot{D}_6-1} \left( \log \frac{1}{x} \right) ^{\ddot{D}_3 \log \frac{1}{x}+\ddot{D}_7} 
\\
\times \ddot{B}_1 \left[ \log \left( 1+ \frac{Q^2}{\ddot{Q}_0^2}\right) - \frac{\ddot{B}_2}{\ddot{B}_1} \left( \frac{1}{\left( 1+\frac{Q^2}{\ddot{Q}_0^2}\right)}-1\right) \right]
\end{multline}
respectively. Putting the extra conditions on the model parameters as
\begin{eqnarray}
& (1) & \ddot{D}_6-1=0 \nonumber \\
& (2) & \ddot{D}_3 \log \frac{1}{x}+\ddot{D}_7=2
\end{eqnarray}
will give the Froissart like behavior in structure function of Eq. \ref{ddot} a new form :
\begin{equation}
\label{x5}
\ddot{F}_2(x,Q^2)= e^{\ddot{D}_0}\ \ddot{Q}_0^2\ \log ^2 \left( 1/x\right)  \ \ddot{B}_1 \left[ \log \left( 1+ \frac{Q^2}{\ddot{Q}_0^2}\right) - \frac{\ddot{B}_2}{\ddot{B}_1} \left( \frac{1}{\left( 1+\frac{Q^2}{\ddot{Q}_0^2}\right)}-1\right) \right]
\end{equation}
\\

The Froissart bound compatible self-similarity based model 2 Eq. \ref{x5} has now power law growth in $\log Q^2$ to be compared with power law growth of model 1.\\

Now using the HERAPDF1.0 \cite{HERA}, Eq.\ref{x5} is fitted and obtained its phenomenological ranges of validity within:  $1.3\times 10^{-4}\leq x \leq 0.02$ and $6.5 \leq Q^2 \leq 60$ GeV$^2$ and also obtained the model parameters which are given in Table \ref{c7t2}. \\

\begin{figure}[tbp]
\captionsetup[subfigure]{labelformat=empty}
\centering
  \subfloat[]{\includegraphics[width=.3\textwidth]{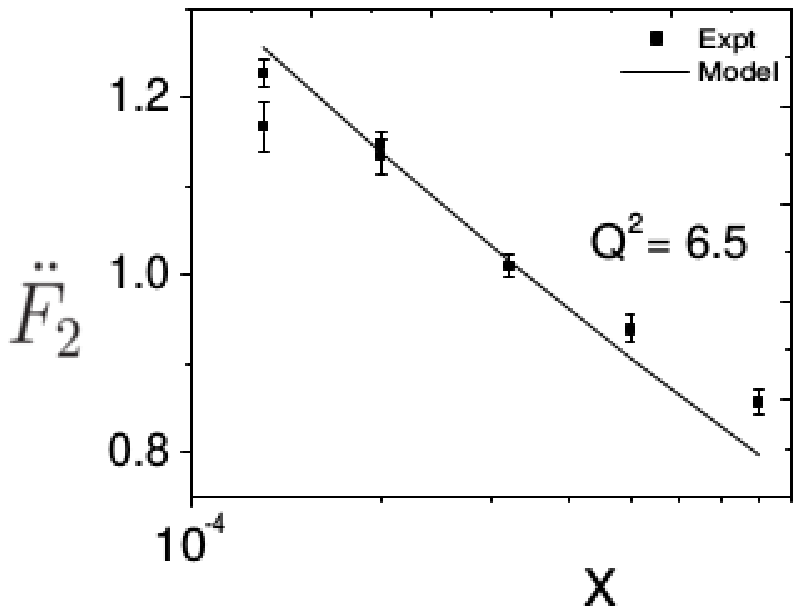}}\quad
 \subfloat[]{\includegraphics[width=.31\textwidth]{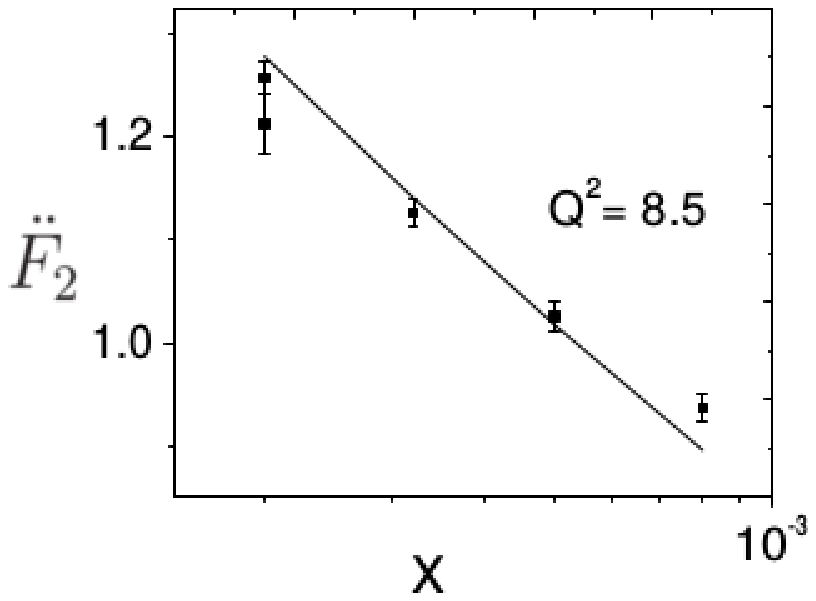}}\quad
\subfloat[]{\includegraphics[width=.3\textwidth]{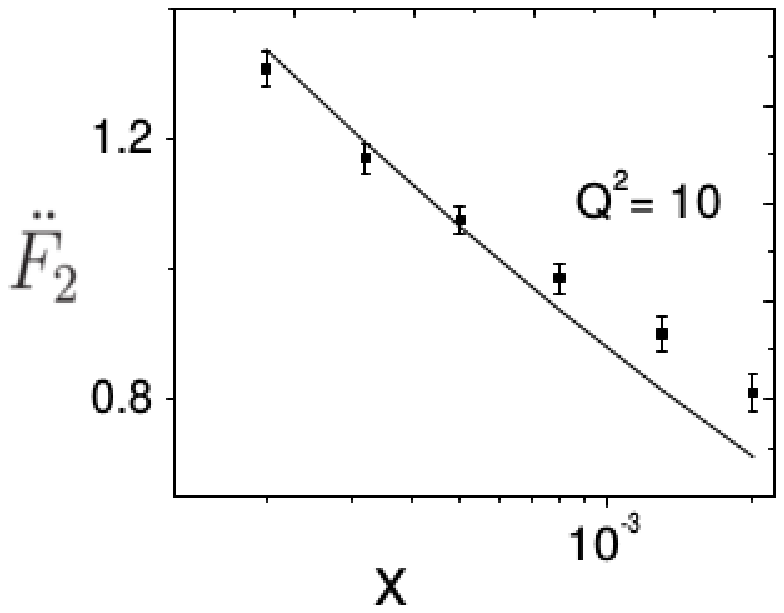}}\quad
 \subfloat[]{\includegraphics[width=.3\textwidth]{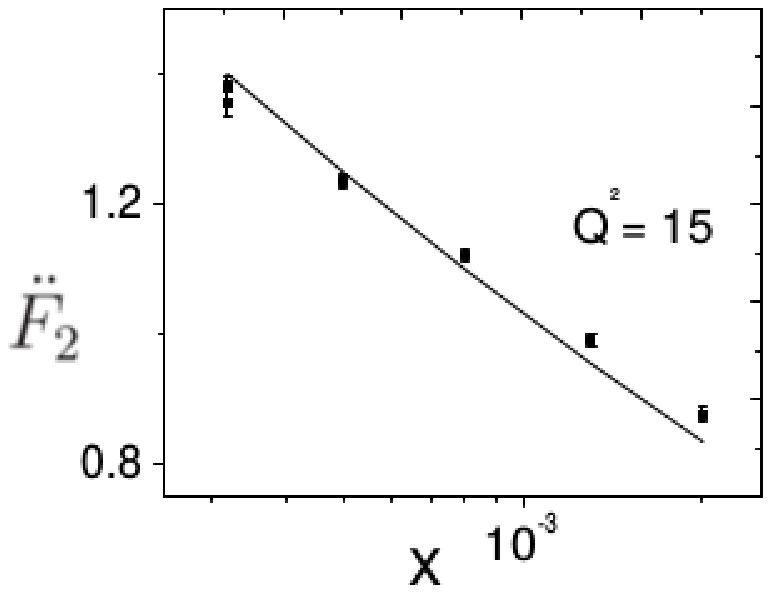}}\quad
 \subfloat[]{\includegraphics[width=.3\textwidth]{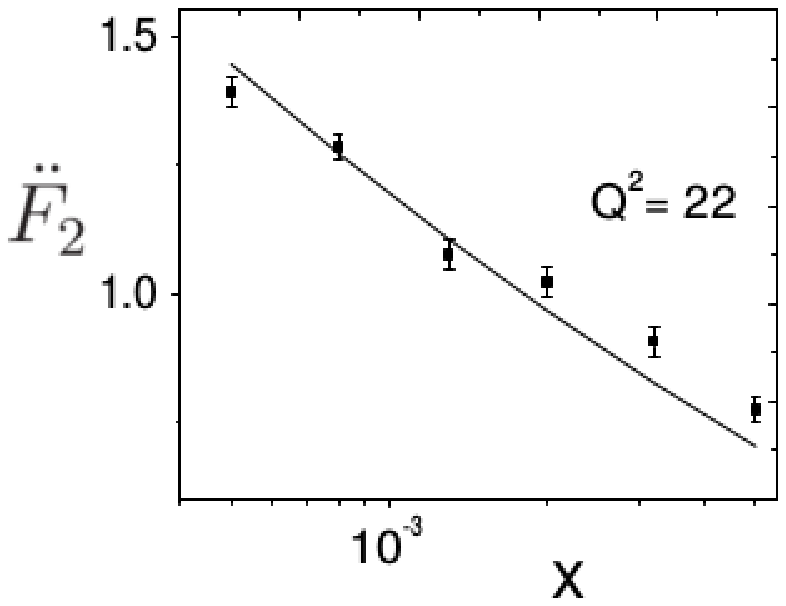}}\quad
  \subfloat[]{\includegraphics[width=.3\textwidth]{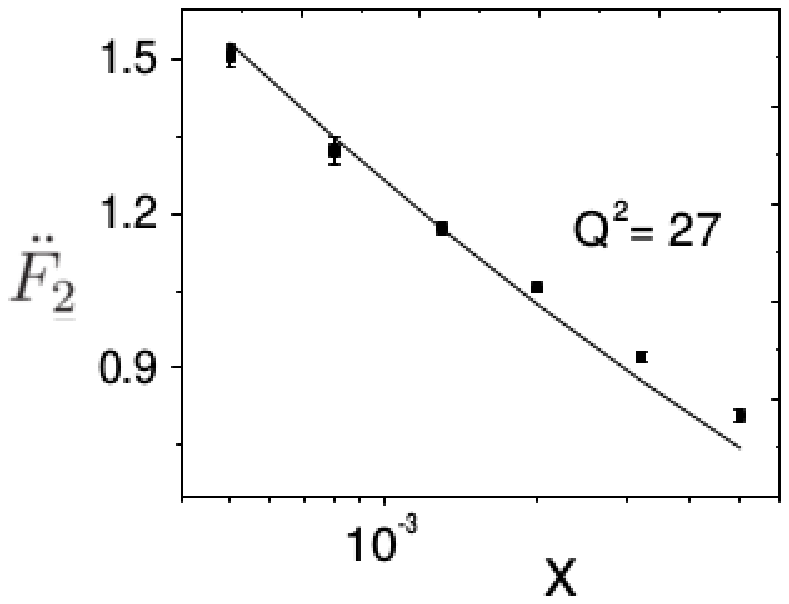}}\quad
\subfloat[]{\includegraphics[width=.3\textwidth]{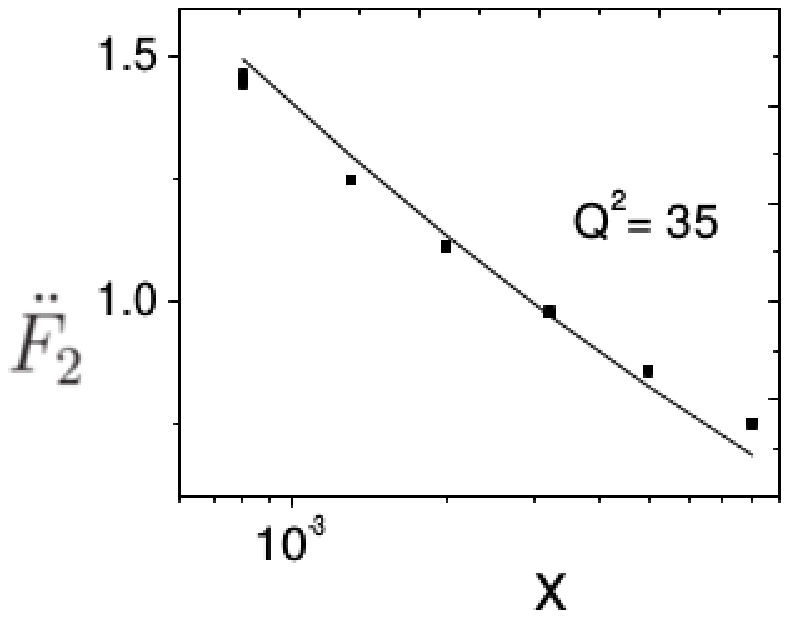}}\quad
\subfloat[]{\includegraphics[width=.3\textwidth]{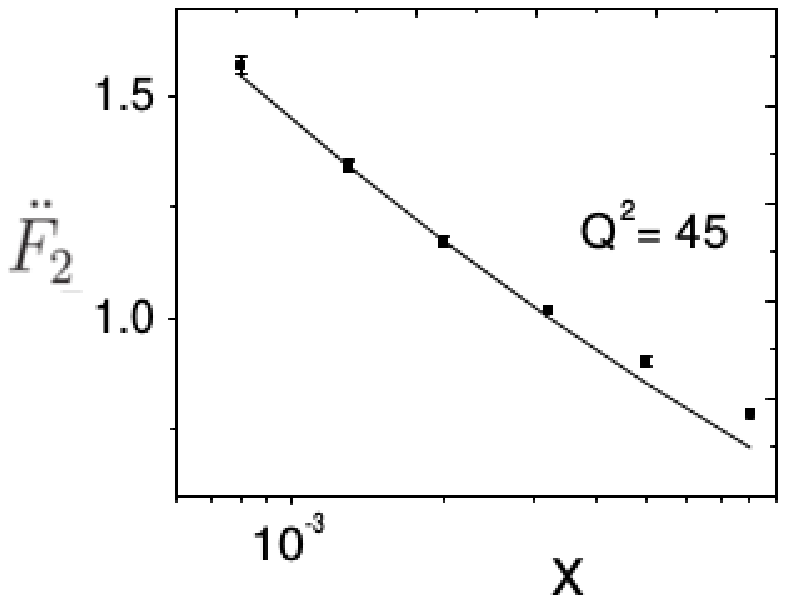}}\quad
 \subfloat[]{\includegraphics[width=.3\textwidth]{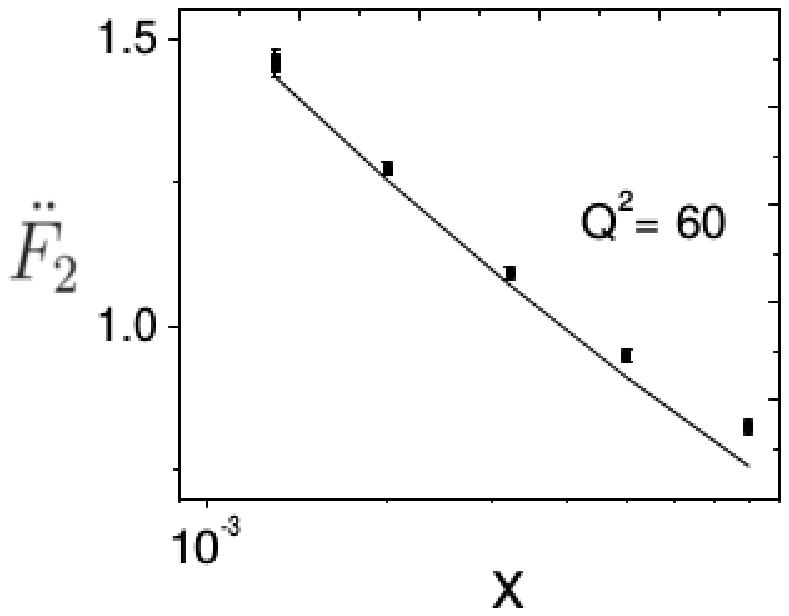}}\quad
 \caption{Comparison of the structure function $\ddot{F}_2$ (Eq \ref{x5}; model 2) as a function of $x$ in bins of $Q^2$ with measured data of $F_2$ from HERAPDF1.0 \cite{HERA}}
 \label{c7F2}
\end{figure}

In Fig \ref{c7F2}, we have shown the graphical representation of $\ddot{F}_2$ with data for a few representative values of $Q^2$. \\

\begin{table}[tbp]
\caption{\label{c7t2}%
\textit{Results of the fit of $\ddot{F}_2$, model 2, Eq.\ref{x5}}}
\begin{ruledtabular}
\begin{tabular}{ccccc}
\textrm{$\ddot{D}_0$}&
\textrm{$\ddot{B}_1$}&
\textrm{$\ddot{B}_2$}&
\textrm{$\ddot{Q}_0^2$(GeV$^2$)}&
\textrm{$\chi^2$/ndf} \\
\colrule
0.00047\tiny${\pm 0.0003}$ & 0.056\tiny${\pm 0.002}$ & 0.672\tiny${\pm 0.02}$ & 0.022\tiny${\pm 0.001}$  &  \\
\end{tabular}
\end{ruledtabular}
\end{table}

\pagebreak
\subsection*{{\large Model 3}}
We now study how far the analytical structure of the models based on self-similarity can come closer to phenomenologically successful model of Block et.al. of Ref. \cite{blo} having Froissart saturation behavior. If the magnification factor $M_2$ is extrapolated to large \textit{x} in a parameter free way $\dfrac{1}{x}\rightarrow \left(\dfrac{1}{x} -1 \right) $, one obtains a set of magnification factors

\begin{eqnarray}
\hat{M}_1 &=& \sum_{j=1}^{n} \frac{B_j}{\left( 1+\frac{Q^2}{\hat{Q}_0^2}\right) ^j}  \quad \quad j=1,2 \nonumber \\
M_2 &=& \dfrac{1}{x}-1 \nonumber \\
M_3 &=& \log \frac{1}{x}
\end{eqnarray}
\\\\
One obtains the following uPDF, PDF and structure function: \\\\
uPDF
\begin{multline}
\breve{f}_i(x,Q^2)= \frac{e^{\breve{D}_0^i}}{M^2}\ (1/x)^{\breve{D}_6}\ (1-x)^{\breve{D}_6} \left( \log \frac{1}{x} \right) ^{\breve{D}_3 \log \frac{1}{x}+\breve{D}_7} 
\\
\times \breve{B}_1 \left[ \frac{1}{\left( 1+ \frac{Q^2}{\breve{Q}_{0}^2}\right)} +\frac{\breve{B}_2}{\breve{B}_1} \frac{1}{\left( 1+ \frac{Q^2}{\breve{Q}_{0}^2}\right)^2}\right]  
\end{multline}
\\\\
Corresponding PDF
\begin{multline}
\breve{q}_i(x,Q^2)=e^{\breve{D}_0^i}\ \breve{Q}_0^2\ (1/x)^{\breve{D}_6} (1-x)^{\breve{D}_6} \left( \log \frac{1}{x} \right) ^{\breve{D}_3 \log \left( \frac{1}{x}-1\right) + \breve{D}_7} 
\\
\times \breve{B}_1 \left[ \log \left( 1+ \frac{Q^2}{\breve{Q}_0^2}\right) - \frac{\breve{B}_2}{\breve{B}_1} \left( \dfrac{1}{\left( 1+ \frac{Q^2}{\breve{Q}_0^2}\right)}-1\right) \right]  
\end{multline}
\\\\
and the structure function
\begin{multline}
\breve{F}_2(x,Q^2)=e^{\breve{D}_0}\ \breve{Q}_0^2\ (1/x)^{\breve{D}_6-1} (1-x)^{\breve{D}_6} \left( \log \frac{1}{x} \right) ^{\breve{D}_3 \log \left( \frac{1}{x}-1\right) +\breve{D}_7} 
\\
\times \breve{B}_1 \left[ \log \left( 1+ \frac{Q^2}{\breve{Q}_0^2}\right) - \frac{\breve{B}_2}{\breve{B}_1} \left( \dfrac{1}{\left( 1+ \frac{Q^2}{\breve{Q}_0^2}\right)}-1\right) \right]  
\end{multline}
\\
Putting the extra conditions 
\begin{eqnarray}
& (1) & \breve{D}_6-1=0 \nonumber \\
& (2) & \breve{D}_3 \log \left( \frac{1}{x}-1\right)+\breve{D}_7=2
\end{eqnarray}
\\
will give the Froissart like behavior in structure function as:\\
\begin{equation}
\label{x6}
\breve{F}_2(x,Q^2)= e^{\breve{D}_0}\ \breve{Q}_0^2\ (1-x) \log ^2 (1/x) \ \times \breve{B}_1 \left[ \log \left( 1+ \frac{Q^2}{\breve{Q}_0^2}\right) - \frac{\breve{B}_2}{\breve{B}_1} \left( \dfrac{1}{\left( 1+ \frac{Q^2}{\breve{Q}_0^2}\right)}-1\right) \right]  
\end{equation}

which has power law growth in $\log Q^2$ rather than in $Q^2$ of model 1.
\\

Using the HERAPDF1.0 \cite{HERA}, Eq.\ref{x6} is fitted and obtained its phenomenological ranges of validity within:  $1.3\times 10^{-4}\leq x \leq 0.02$ and $6.5 \leq Q^2 \leq 120$ GeV$^2$ and also obtained the model parameters which are given in Table \ref{c7t3}.
\\

\begin{figure}[!tbp]
\captionsetup[subfigure]{labelformat=empty}
\centering
  \subfloat[]{\includegraphics[width=.3\textwidth]{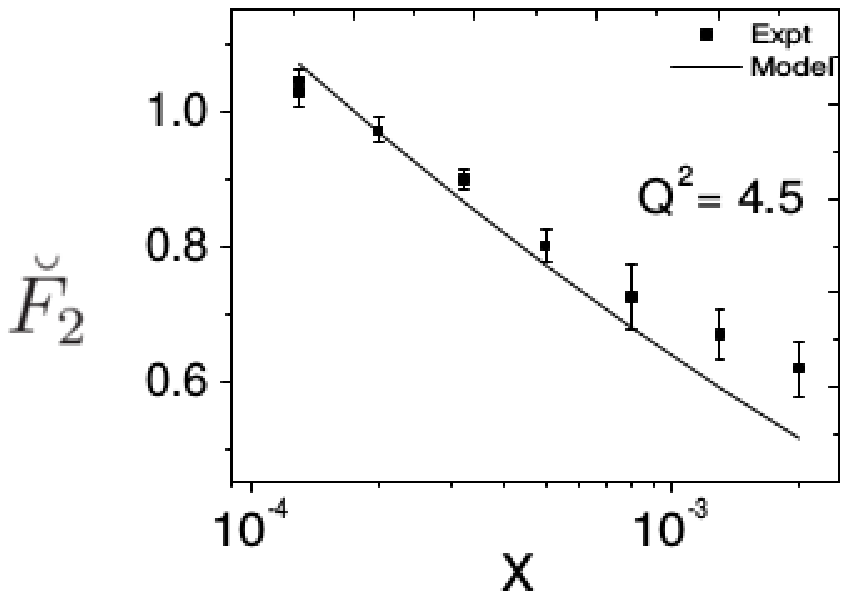}}\quad
 \subfloat[]{\includegraphics[width=.31\textwidth]{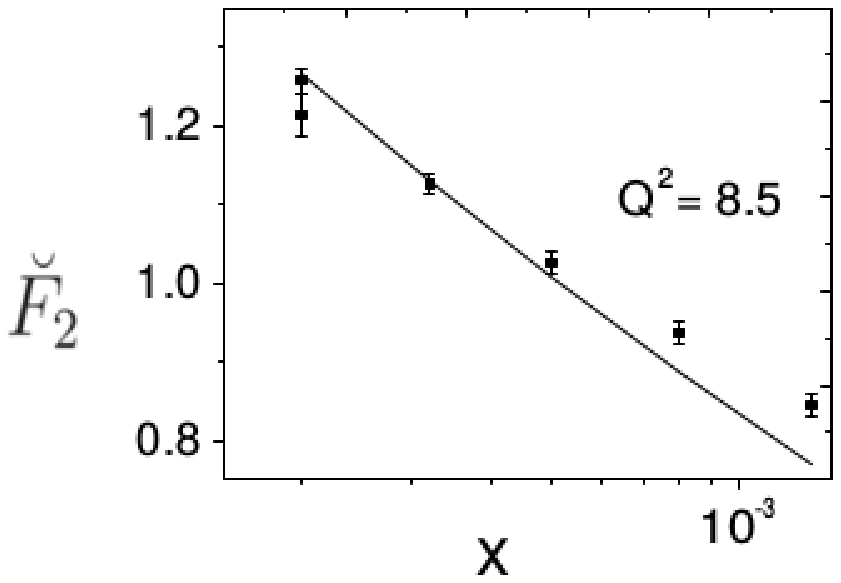}}\quad
\subfloat[]{\includegraphics[width=.3\textwidth]{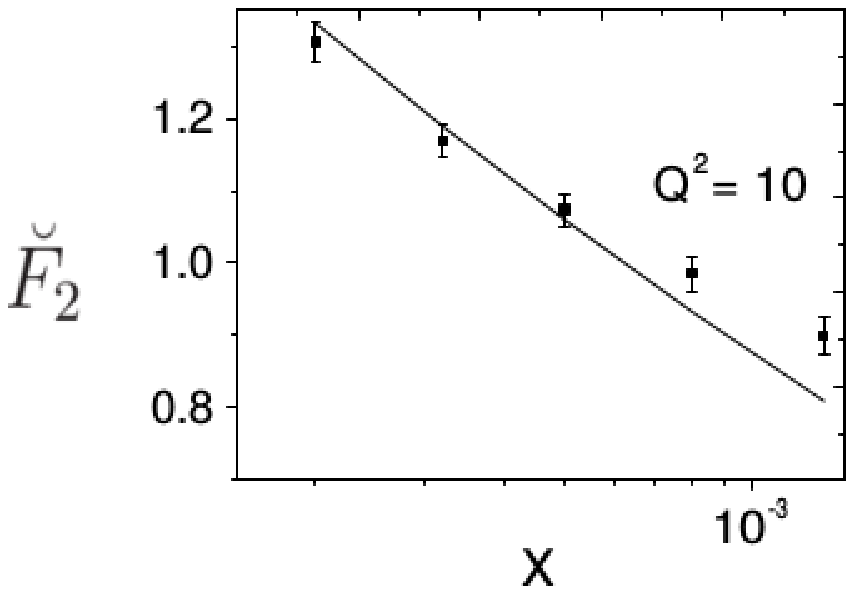}}\quad
 \subfloat[]{\includegraphics[width=.3\textwidth]{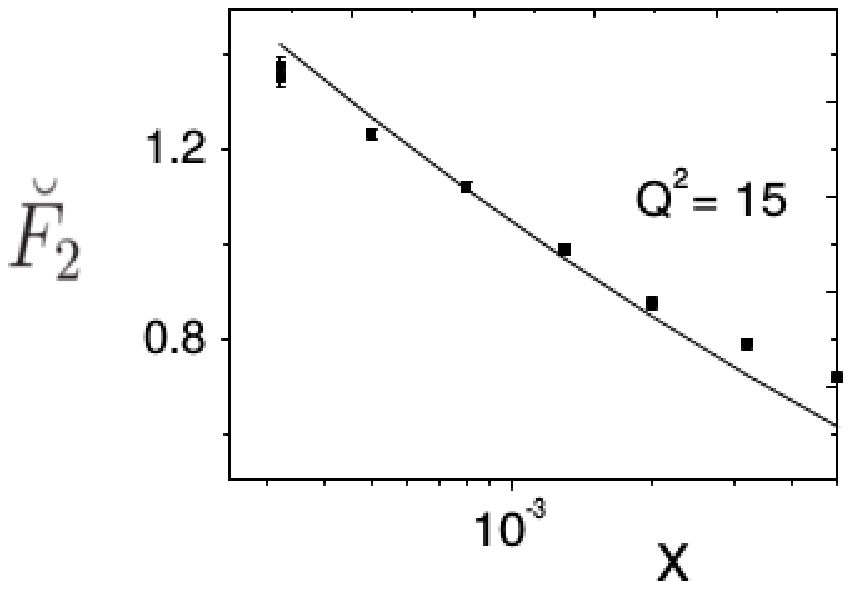}}\quad
 \subfloat[]{\includegraphics[width=.3\textwidth]{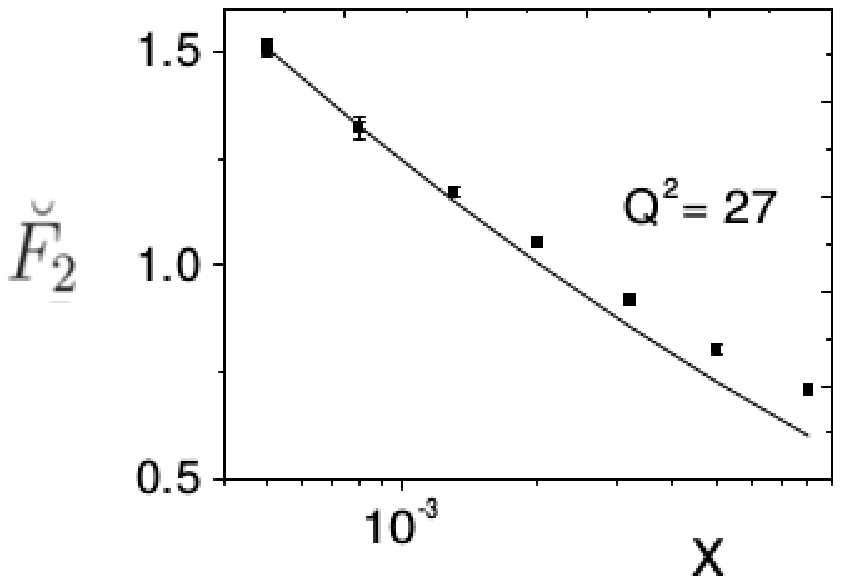}}\quad
  \subfloat[]{\includegraphics[width=.3\textwidth]{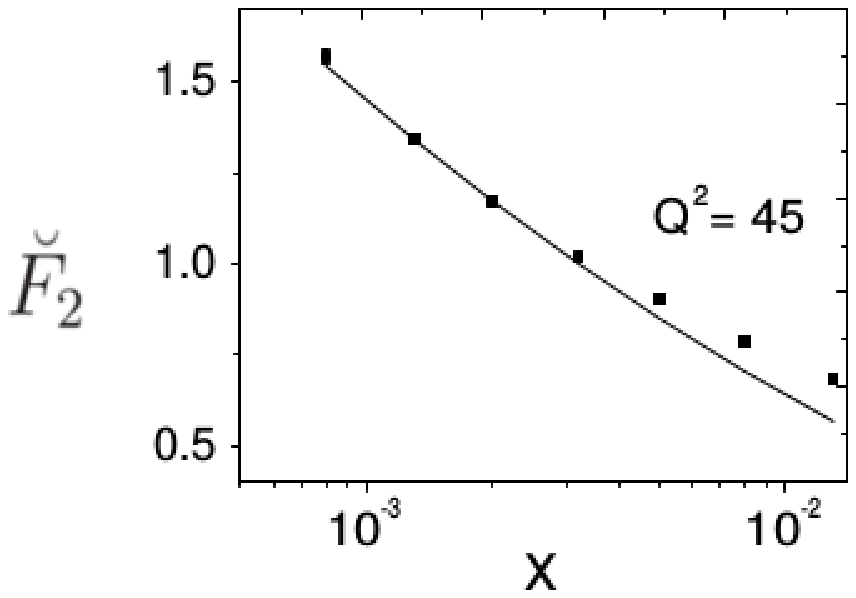}}\quad
\subfloat[]{\includegraphics[width=.3\textwidth]{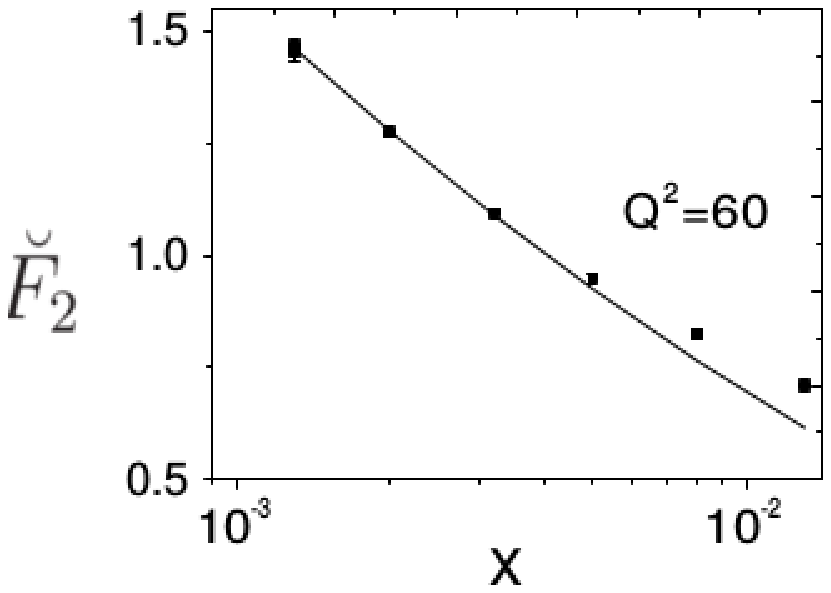}}\quad
\subfloat[]{\includegraphics[width=.3\textwidth]{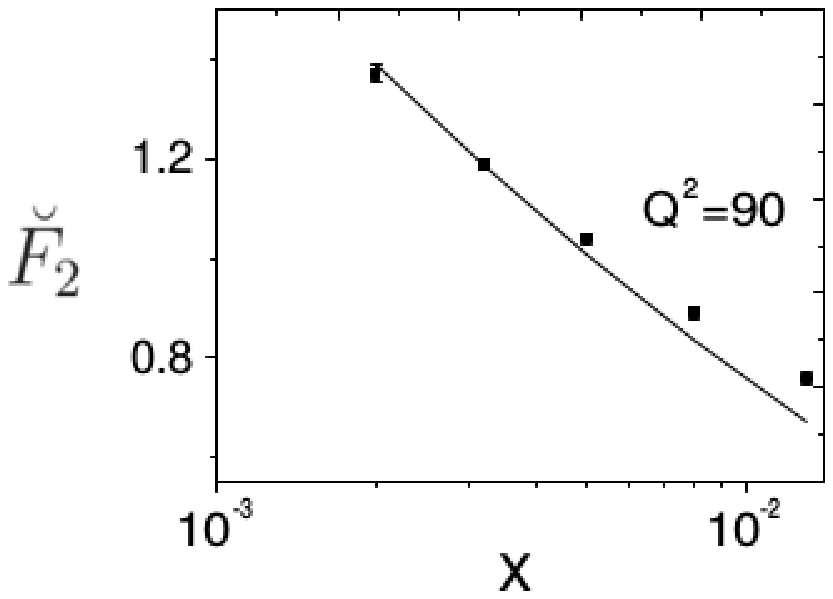}}\quad
 \subfloat[]{\includegraphics[width=.3\textwidth]{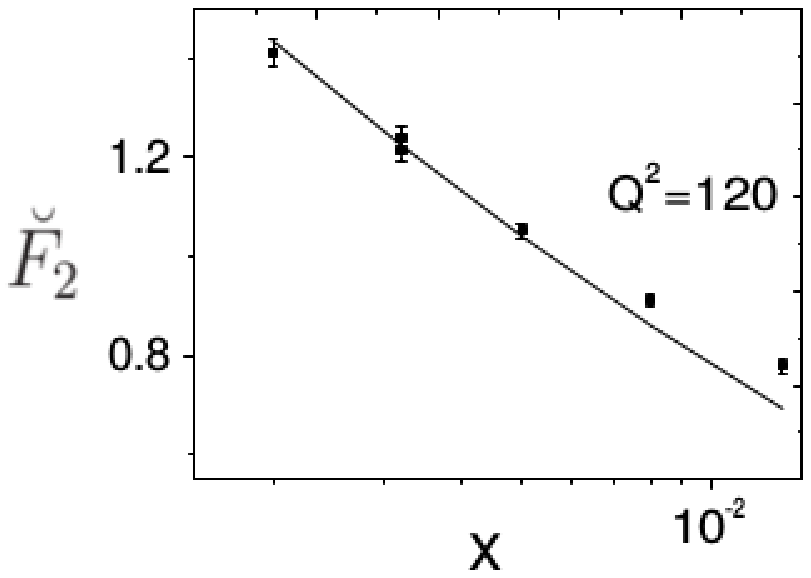}}\quad
 \caption{Comparison of the structure function $\breve{F}_2$ (Eq \ref{x6}; model 3) as a function of $x$ in bins of $Q^2$ with measured data of $F_2$ from HERAPDF1.0 \cite{HERA}}
 \label{c7F3}
\end{figure}

In Fig. \ref{c7F3}, we have shown the graphical representation of $\breve{F}_2$ with data for a few representative values of $Q^2$. 

\begin{table}[!bp]
\caption{\label{c7t3}%
\textit{Results of the fit of $\breve{F}_2$, model 3, Eq.\ref{x6}}}
\begin{ruledtabular}
\begin{tabular}{ccccc}
\textrm{$\breve{D}_0$}&
\textrm{$\breve{B}_1$}&
\textrm{$\breve{B}_2$}&
\textrm{$\breve{Q}_0^2$(GeV$^2$)}&
\textrm{$\chi^2$/ndf} \\
\colrule
0.006\tiny${\pm 0.0005}$ & 0.032\tiny${\pm 0.0005}$ & 0.309\tiny${\pm 0.009}$ & 0.048\tiny${\pm 0.001}$  & 0.25 \\
\end{tabular}
\end{ruledtabular}
\end{table}

\pagebreak
\subsection*{{\large Model 4}}
If the third magnification factor is also large-\textit{x} extrapolated: $\log \dfrac{1}{x} \rightarrow \log \left(\dfrac{1}{x} -1 \right)$ i.e
\begin{eqnarray}
\hat{M}_1 &=& \sum_{j=1}^{n} \frac{B_j}{\left( 1+\frac{Q^2}{\hat{Q}_0^2}\right) ^j}  \quad \quad j=1,2 \nonumber \\
M_2 &=& \dfrac{1}{x}-1 \nonumber \\
M_3 &=& \log \left( \frac{1}{x}-1 \right) 
\end{eqnarray}
\\
the corresponding uPDF PDF and structure function becomes:
\\
uPDF
\begin{multline}
\breve{f}'_i(x,Q^2)= \frac{e^{\breve{D}_0'^i}}{M^2}\ (1/x)^{\breve{D}'_6}\ (1-x)^{\breve{D}'_6} \left( \log \frac{1-x}{x} \right) ^{\breve{D}'_3 \log \left( \frac{1}{x}-1\right) +\breve{D}'_7} 
\\
\times \breve{B}'_1 \left[ \frac{1}{\left( 1+ \frac{Q^2}{\breve{Q}_{0}'^2}\right)} +\frac{\breve{B}'_2}{\breve{B}'_1} \frac{1}{\left( 1+ \frac{Q^2}{\breve{Q}_{0}'^2}\right)^2}\right]  
\end{multline}
Corresponding PDF
\begin{multline}
\breve{q}'_i(x,Q^2)=e^{\breve{D}_0'^i}\ \breve{Q}_0'^2\ (1/x)^{\breve{D}'_6} (1-x)^{\breve{D}'_6} \left( \log \frac{1-x}{x} \right) ^{\breve{D}'_3 \log \left( \frac{1}{x}-1\right) +\breve{D}'_7} 
\\
\times \breve{B}'_1 \left[ \log \left( 1+ \frac{Q^2}{\breve{Q}_0'^2}\right)- \frac{\breve{B}'_2}{\breve{B}'_1} \left( \dfrac{1}{\left( 1+ \frac{Q^2}{\breve{Q}_0'^2}\right)}-1\right) \right]  
\end{multline}
and the structure function
\begin{multline}
\breve{F}'_2(x,Q^2)=e^{\breve{D}'_0}\ \breve{Q}_0'^2\ (1/x)^{\breve{D}'_6-1} (1-x)^{\breve{D}'_6} \left( \log \frac{1-x}{x} \right) ^{\breve{D}'_3 \log \left( \frac{1}{x}-1\right) +\breve{D}'_7} 
\\
\times \breve{B}'_1 \left[ \log \left( 1+ \frac{Q^2}{\breve{Q}_0'^2}\right) - \frac{\breve{B}'_2}{\breve{B}'_1} \left( \dfrac{1}{\left( 1+ \frac{Q^2}{\breve{Q}_0'^2}\right)}-1\right) \right]   
\end{multline}
\\
Putting the extra conditions
\begin{eqnarray}
& (1) & \breve{D}'_6-1=0 \nonumber \\
& (2) & \breve{D}'_3 \log \left( \frac{1}{x}-1\right)+\breve{D}'_7=2
\end{eqnarray}
can show the Froissart like behavior in structure as: 
\\
\begin{equation}
\label{x7}
\breve{F}'_2(x,Q^2)= e^{\breve{D}'_0}\ \breve{Q}_0'^2\ (1-x)\ \log ^2 \left( \frac{1-x}{x}\right)  \ \breve{B}'_1 \left[ \log \left( 1+ \frac{Q^2}{\breve{Q}_0'^2}\right) - \frac{\breve{B}'_2}{\breve{B}'_1} \left( \dfrac{1}{\left( 1+ \frac{Q^2}{\breve{Q}_0'^2}\right)}-1\right) \right]  
\end{equation}

Eq. \ref{x7} of model 4 based on self-similarity and Froissart bound compatibility is closest to the phenomenologically successful model suggested by Block, Durand, Ha and McKay which has a wide range of phenomenological validity in $Q^2$: $0.11\leq Q^2\leq 1200$ GeV$^2$ for small $x\leq x_p=0.11$ \cite{blooo} together with a Froissart Saturation like behavior \cite{fe}.\\

\begin{table}[!bp]
\caption{\label{c7t4}%
\textit{Results of the fit of $\breve{F}'_2$, model 4, Eq.\ref{x7}}}
\begin{ruledtabular}
\begin{tabular}{ccccc}
\textrm{$\breve{D}'_0$}&
\textrm{$\breve{B}'_1$}&
\textrm{$\breve{B}'_2$}&
\textrm{$\breve{Q}_0'^2$(GeV$^2$)}&
\textrm{$\chi^2$/ndf} \\
\colrule
0.008\tiny${\pm 0.001}$ & 0.034\tiny${\pm 0.0008}$ & 0.251\tiny${\pm 0.01}$ & 0.057\tiny${\pm 0.005}$  & 0.26 \\
\end{tabular}
\end{ruledtabular}
\end{table}

\begin{figure}[!bp]
\captionsetup[subfigure]{labelformat=empty}
\centering
  \subfloat[]{\includegraphics[width=.3\textwidth]{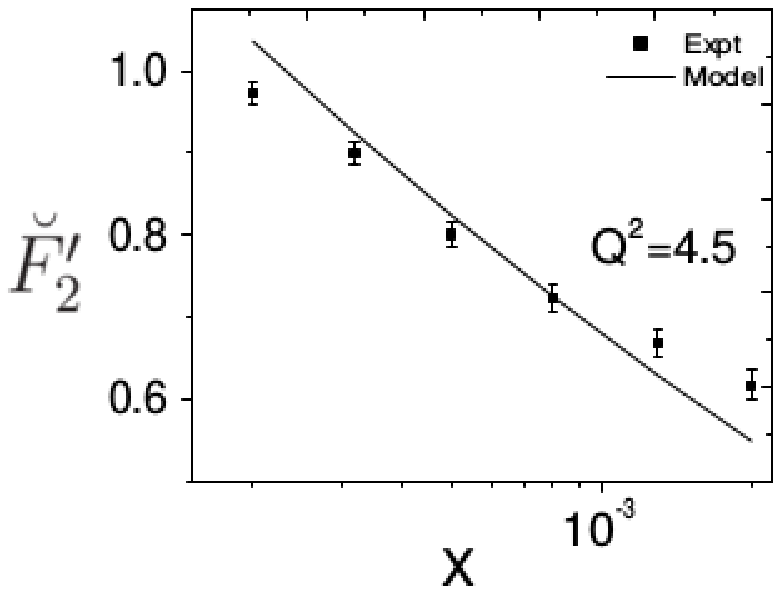}}\quad
 \subfloat[]{\includegraphics[width=.3\textwidth]{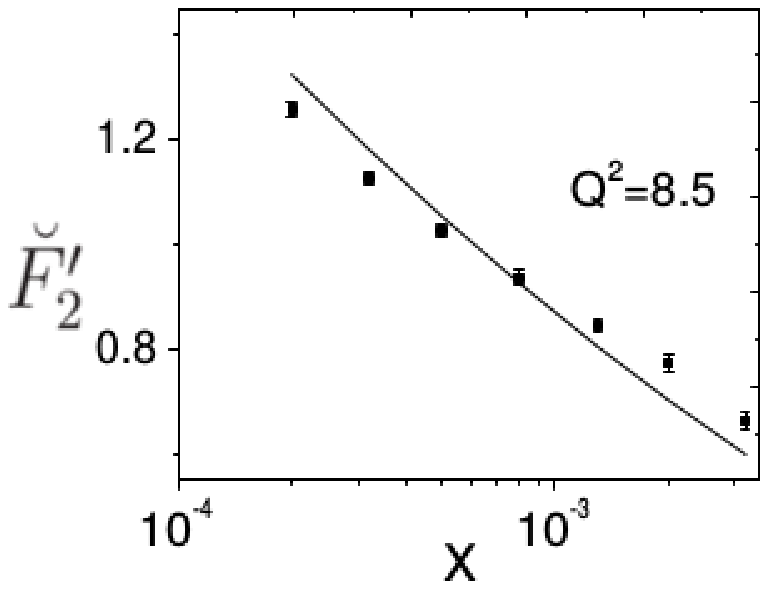}}\quad
\subfloat[]{\includegraphics[width=.3\textwidth]{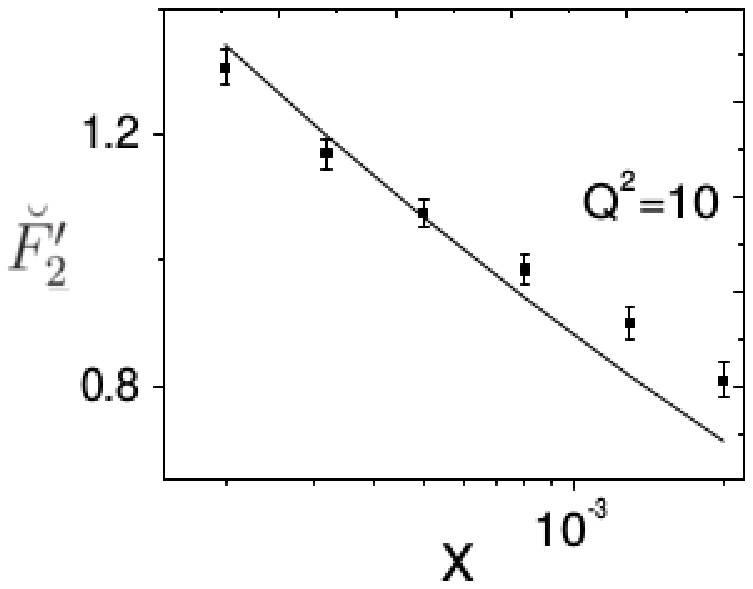}}\quad
 \subfloat[]{\includegraphics[width=.3\textwidth]{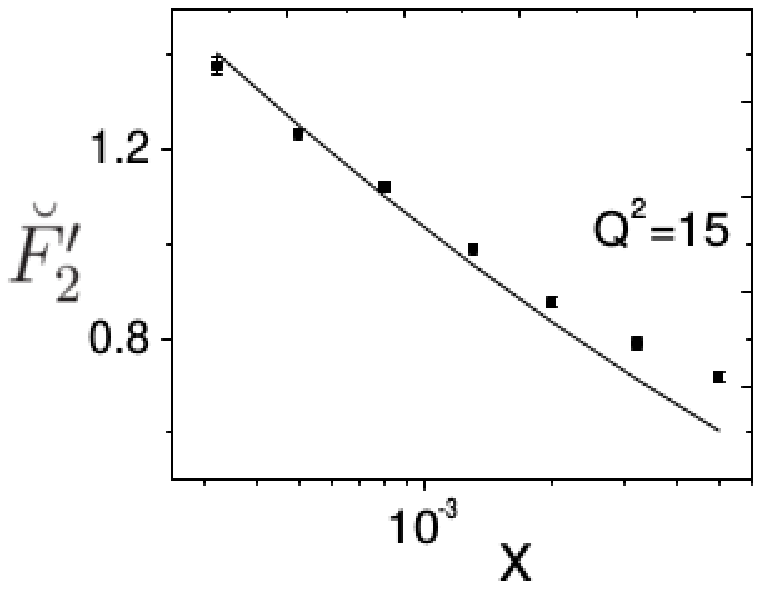}}\quad
 \subfloat[]{\includegraphics[width=.3\textwidth]{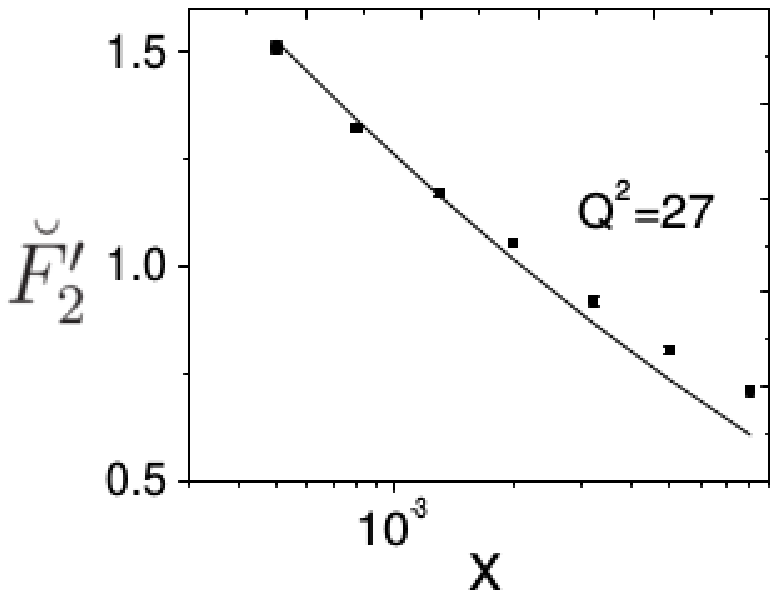}}\quad
  \subfloat[]{\includegraphics[width=.3\textwidth]{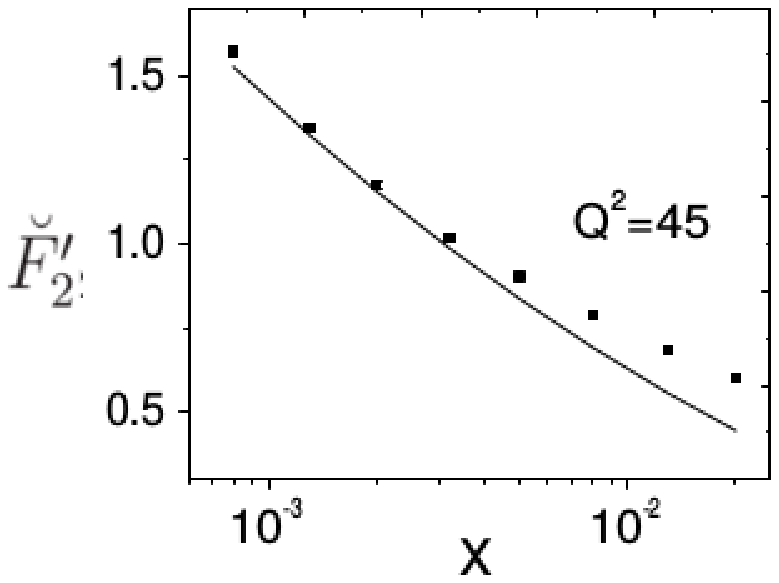}}\quad
\subfloat[]{\includegraphics[width=.31\textwidth]{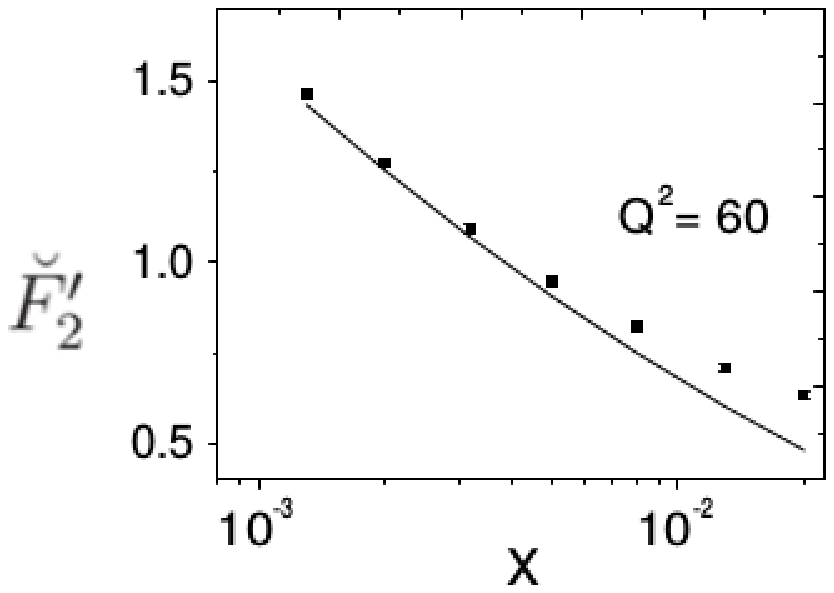}}\quad
\subfloat[]{\includegraphics[width=.3\textwidth]{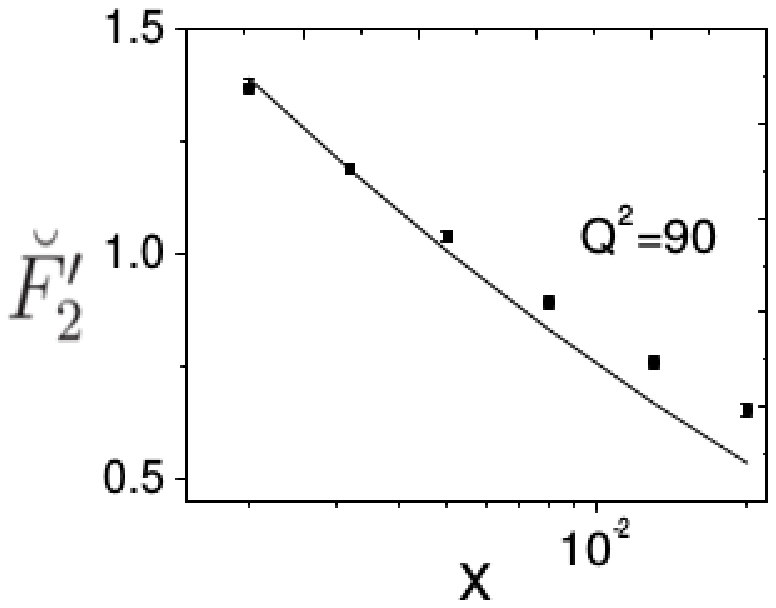}}\quad
 \subfloat[]{\includegraphics[width=.3\textwidth]{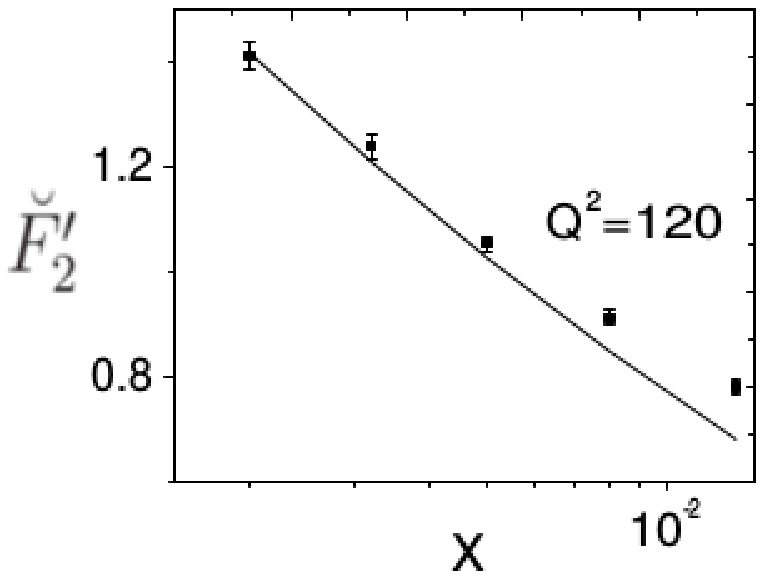}}\quad
 \caption{Comparison of the structure function $\breve{F}'_2$ (Eq. \ref{x7}; model 4) as a function of $x$ in bins of $Q^2$ with measured data of $F_2$ from HERAPDF1.0 \cite{HERA}}
 \label{c7F4}
\end{figure}

Using the HERAPDF1.0 \cite{HERA}, Eq. \ref{x7} is fitted and obtained its phenomenological ranges of validity within:  $1.3\times 10^{-4}\leq x \leq 0.02$ and $6.5 \leq Q^2 \leq 120$ GeV$^2$ with the obtained model parameters which are given in Table \ref{c7t4}. In Fig. \ref{c7F4}, we have shown the graphical representation of $\breve{F}'_2$ with data for a few representative values of $Q^2$.

\section*{{\large Model 5}}
The expression for $F_2^p(x,Q^2)$ of Ref. \cite{blo} is:
\begin{equation}
\label{c6E31}
F_2^p(x,Q^2) = (1-x)\left\lbrace \frac{F_p}{1-x_p}+A(Q^2)\ln\frac{x_p (1-x)}{x (1-x_p)}+ B(Q^2)\ln^2\frac{x_p (1-x)}{x (1-x_p)} \right\rbrace 
\end{equation}
Where,
\begin{eqnarray}
A(Q^2) = a_0 + a_1 \ln Q^2 + a_2 \ln^2 Q^2 \nonumber \\
B(Q^2) = b_0 + b_1 \ln Q^2 + b_2 \ln^2 Q^2
\end{eqnarray}
\\
and the parameters fitted from deep inelastic scattering data \cite{blo} are
\begin{eqnarray}
x \leqslant x_p = 0.11 \ \ {\text a\text n\text d} \ \ F_p = 0.413 \pm 0.003 \ ,  \\ \nonumber
\end{eqnarray}
\begin{eqnarray}
\label{c6Ee}
a_0 &=& -8.471\times 10^{-2}\pm 2.62\times 10^{-3} \ , \nonumber \\
a_1 &=& 4.190\times 10^{-2}\pm 1.56\times 10^{-3} \ , \nonumber \\
a_2 &=& -3.976\times 10^{-3}\pm 2.13\times 10^{-4} \ , \nonumber \\
b_0 &=& 1.292\times 10^{-2}\pm 3.62\times 10^{-4} \ , \nonumber \\
b_1 &=& 2.473\times 10^{-4}\pm 2.46\times 10^{-4} \ , \nonumber \\
b_2 &=& 1.642\times 10^{-3}\pm 5.52\times 10^{-5} \ . \
\end{eqnarray}
\\
More recently, expression of Eq. \ref{c6E31} was used as an input at $Q^2=4.5$ GeV$^2$ in DGLAP evolution equations in leading order (LO) and obtained a wider phenomenological $Q^2$-range upto $Q^2\leq 3000$ GeV$^2$  using more recent HERA data \cite{HERA}.\\

One can write the Eq. \ref{c6E31} in a more simplified version:
\begin{multline}
\label{block1}
F_2^{\text p} \sim (1-x) \left[  \ C + a_0' \ln \frac{1-x}{x}+ a_1' \ln Q^2 \ln \frac{1-x}{x} + a_2' \ln^2 Q^2 \ln \frac{1-x}{x}  \right.\
\\
+ \left. b_0'\ln^2 \frac{1-x}{x} + b_1'\ln Q^2 \ln^2 \frac{1-x}{x} + b_2'\ln^2 Q^2 \ln^2 \frac{1-x}{x} \ \right] 
\end{multline}
where \\
$C= \dfrac{F_p}{1-x_p}$ , \ $a_0' = a_0 \dfrac{x_p}{1-x_p}$ , \  $a_1' = a_1 \dfrac{x_p}{1-x_p}$ , \ $a_2' = a_2 \dfrac{x_p}{1-x_p}$ , \\ \\ \ $b_0' = b_0 \dfrac{x_p}{1-x_p}$ , \ $b_1' = b_1 \dfrac{x_p}{1-x_p}$ , \ $b_2' = b_2 \dfrac{x_p}{1-x_p}$ \\

The relative success of the model 5 over model 4 is that while the model 5 has got additional multiplicative terms like $\ln^2 Q^2$ and $\ln^2 Q^2$ with leading and non-leading terms of the order of $\log \dfrac{1}{x}$ with multiplicative factors $\ln^2 Q^2$ and $\ln^2 Q^2$, the present self-similarity based model 4 do not have such additional leading and non-leading additive/multiplicative terms.

The analysis indicates that Froissart Saturation like behavior is possible in the self-similarity based models of proton structure function pursued by us, provided the magnification factors increases from 2 to 3, with an additional magnification factor $\log \frac{1}{x}$. We have studied the models having either power law growth in $Q^2$ (model 1) or in $\log Q^2$ (models 2-4). Each of them has $\log^2 \frac{1}{x}$ rise rather than $\left( \frac{1}{x}\right)^m$, where $m>0$ in previous version of the models \cite{bs1}. The present analysis however with self-similarity based models seems to prefer a faster power law growth in $\sim \left( \frac{1}{x}\right)^m$ rather than $\sim  \log^2 \frac{1}{x}$ when compared with data.

\section{Summary}
\label{summ}
In this analysis, we have obtained the Froissart saturated forms of structure function $F_2^p$ having both the power law rise in $Q^2$ and $\log Q^2$. It needs at least three magnification factors not two as compared to earlier work in Ref. \cite{dkc}. The ranges of validity for four different self-similarity based models with three magnification factors are: 
\begin{eqnarray}
& \text{Model 2}; \ {\text E\text q}. \ref{x4} \ : & 1.3\times 10^{-4}\leq x \leq 0.02 \ ; \ 6.5 \leq Q^2 \leq 120 \ {\text G\text e\text V^2} \nonumber \\
& \text{Model 3}; \ {\text E\text q}. \ref{x5} \ : & 1.3\times 10^{-4}\leq x \leq 0.02 \ ; \ 6.5 \leq Q^2 \leq 60 \ {\text G\text e\text V^2} \nonumber \\
& \text{Model 4}; \ {\text E\text q}. \ref{x6} \ : & 1.3\times 10^{-4}\leq x \leq 0.02 \ ; \ 6.5 \leq Q^2 \leq 120 \ {\text G\text e\text V^2} \nonumber \\
& \text{Model 5}; \ {\text E\text q}. \ref{x7} \ : & 1.3\times 10^{-4}\leq x \leq 0.02 \ ; \ 6.5 \leq Q^2 \leq 120 \ {\text G\text e\text V^2} \nonumber
\end{eqnarray}
to be compared with the models of Ref. \cite{bs1} based on self-similarity without Froissart saturation condition:
\begin{eqnarray}
& {\text E\text q}. 34 \ : & 2\times10^{-5}\leq x\leq 0.4 \ ; \ 1.2 \leq Q^2 \leq 800  \ {\text G\text e\text V^2} \nonumber \\
& {\text E\text q}. 44 \ : & 2\times10^{-5}\leq x\leq 0.4 \ ; \ 1.2 \leq Q^2 \leq 1200 \ {\text G\text e\text V^2} \nonumber
\end{eqnarray}
Thus the above analysis shows the Froissart saturated self-similarity based structure function has smaller validity ranges as compared to that of structure function without Froissart condition having power law growth in $\frac{1}{x}$ and $\log Q^2$.

So our inference is that perhaps the present HERA data has not reached its asymptotic regime to have a Froissart saturation like behavior if self-similarity is assumed to be a symmetry of the structure function. 
\\

Let us end this section with the theoretical limitation of the present work. 

Although fractality in hadron-hadron and electron-positron interactions has been well established experimentally \cite{7} , self-similarity itself is not a general property of QCD and is not yet established, either theoretically or experimentally. In this work, we have merely used the notion of self-similarity to parametrize PDFs/structure functions as a generalization of the method suggested in Ref. \cite{Last} and reported by us recently in \cite{bs1,bsc} and then study how additional  conditions among the model parameters are needed to make them compatible with Froissart Bound as well. Besides $\log^2 (1/x)$ behavior, the presented models have also   power law rise in $\log Q^2$ rather than in $Q^2$ and are closer to QCD expectation. However, in no way, the constructed fractal inspired  models are comparable to those  based on QCD. Modern analysis of PDFs in perturbative QCD are carried out upto  Next-to-Next-to-leading order (NNLO) \cite{nnlo,nnlo1} with and without Froissart saturation using standard QCD evolution equation and corresponding calculable splitting functions in several orders of strong coupling constant and compare with QCD predictions. Instead, the present work is carried out only at the level of a parton model. In this way, the models merely parametrize the input parton distributions and their evolution in a self-similarity based compact form, which contains both perturbative and non-perturbative aspects of a formal theory, valid in a finite $x-Q^2$ range of data. It presumably implies that while self-similarity has not yet been proved to be a general feature of strong interactions, under specific conditions, experimental data can be interpreted with this notion as has been shown in the present paper. However, to prove it from the first principle is beyond its scope.

\section*{Acknowledgment}
The formalism of the present paper was initiated when one of us (DKC) visited the Rudolf Peirels Center of Theoretical Physics, University of Oxford. He thanks Professor
Subir Sarkar and Professor Amanda Cooper-Sarkar for useful discussion. We also thank Dr. Kushal Kalita for helpful discussions. One of the authors (BS) acknowledges the UGC-RFSMS for financial support.


\end{document}